# Systematic behavior of the fusion barrier parameters for heavy ion pairs

**By**

**Abdulghany Reda Abdulghany Ahmed**

**A Thesis Submitted to**

**Faculty of Science**

**In Partial Fulfillment of the Requirements for the degree of**

**Master of Science**

**(Theoretical physics)**

**Physics Department**

**Faculty of Science**

**Cairo University**

**(2010)**

**APROVAL SHEET FOR SUBMISSION**

Thesis Title

# Systematic behavior of the fusion barrier parameters for heavy ion pairs

Name of candidate

**Abdulghany Reda Abdulghany Ahmed**

**This thesis has been approved for submission by the supervisors:**

**1- Prof. Dr. Mahmoud Yahia Ismail**
Physics Department - Faculty of Science - Cairo University

Signature:

**2- Dr. Ali Yahia Ellithi**
Physics Department - Faculty of Science- Cairo University

Signature:

**Prof. Dr. Gamal Abd-Elnaser Madbouly**
Chairman of Physics Department
Faculty of Science - Cairo University

Signature:

# ABSTRACT

The nucleus-nucleus potential is calculated in the frame work of the double folding model (DFM) to obtain the Coulomb barrier parameters (barrier position and height), starting from M3Y-Reid nucleon-nucleon interaction and realistic nuclear matter distribution. The systematic behavior of the barrier parameters with mass numbers, charges, and radii of interacting nuclei is studied. The relation between the barrier height and radius is also discussed. The systematic behavior of the barrier parameters is presented in the form of simple analytical formulae, which can be used to calculate the barrier position and height directly, and show which factors can affect them.

The potentials obtained from DFM are used to derive a universal function of the nuclear proximity potential which is useful for barrier calculations for heavy ion reactions. The obtained universal function reproduces the barrier parameters within less than 2% deviation from the values obtained using DFM for heavy and super heavy ion reactions. Reactions involving α-particle are studied individually, and another form of the universal function is presented.

# Acknowledgements

First of all, I would like to thank my supervisor Prof. Dr. Mahmoud Yahia Ismail, who guided me to complete successfully my thesis, and who also have kindly and meticulously undertaken the tedious work of proofreading my thesis, I would express my appreciation to him not only for his professional advice and guidance, but also for his enthusiasm and encouragement. I would also like to thank Dr. Ali Yahia Ellithi, my second supervisor, who helped me on my simulation work, and whose comments and questions have helped me deepen my understanding about many parts of the research.

I am indebted to many of my colleagues, who helped me through the year by their friendships and by giving helpful comments, and supported me all the time.

I would also like to express my gratitude to my wife Marwa, without her support, this thesis would have never come to completion.

Abdulghuny, 2010.

# Table of Contents

## List of Tables

## List of Figures

# Preface

In the last few decades, the study of nuclear reactions became one of the most interesting fields of physics. The main focus was on production of energy; however, nuclear studies affect many vital fields, such like medicine, biology, archeology and militarism. As the building of accelerators developed the scientific ambition developed, and many studies were performed on the synthesis of new super heavy elements (SHE). This opened up a new field of research termed as heavy-ion collision physics. Most of our information is obtained from studies made on stable nuclei for the simple reason that they are far easier to handle in the laboratory. This is a very special group among all the possible ones that can be formed. Synthesis of SHE is a great challenging topic, not to get the SHE itself, but to get a detailed picture about the shell stabilization and structure effects. This also gives a new tool to test the nuclear theories, or even develop them.

The building blocks of nuclei are neutrons and protons, two quantum states of the same particle, the nucleon. Both gravitational and electromagnetic forces are infinite in range and their interaction strengths diminish with the square of the distance of separation. Clearly, nuclear force cannot follow the same radial dependence. The nuclear force has a very short range, not much beyond the confine of the nucleus itself. In 1935, Yukawa proposed that the force between nucleons arises from meson exchange. This was the start of the concept of field quantum as the mediator of fundamental forces. The reason that nuclear force has a finite range comes from the nonzero rest mass of the mesons exchanged. For the nucleons inside a nucleus, nuclear force is far stronger than that due to electromagnetic interaction. This force keeps the protons and the neutrons bound to the nucleus, and it makes the nuclear reactions possible.

The interaction between two nuclei is governed by the repulsive Coulomb potential and the attractive nuclear potential, which in combination form the potential barrier; this barrier has to be penetrated for fusion to occur. Understanding the physics of fusion of heavy ions is still a central topic of research in nuclear physics. For this purpose many studies of fusion barrier have been done. Well knowledge of fusion barrier parameters (barrier height and barrier position) gives a great idea about the process of fusion and tells us about the conditions needed to get the wanted result. In chapter (1) of the present thesis we review the effect of different factors such like, masses, charges, diffuseness of nuclear matter, and radii of the interacting pair on the barrier parameters. We introduce the behavior of the fusion barrier parameters in the form of simple analytical formulae, not only to get a simple method to predict the values of barrier parameters, but also to show which factors can affect them. In chapter (2) we introduce a method to calculate nuclear potential around the barrier position. We use the advantages of two different models, the first is the “double folding model” which characterized by its great validity in the tail region (around the barrier position), and the second is the “proximity model” which characterized by the accessibility in the calculation of nuclear interaction. We used the results of detailed calculations through the double folding model to introduce a new shape of the universal function useful in barrier calculations for heavy ion reactions. We also study the reactions involve α-particle individually, because of its odd characteristics, and we introduce a universal function useful for reactions involve α-particle and α-decay.

# *Chapter 1*

# *Systematic behavior of the fusion barrier parameters using the double folding model*

# Chapter 1

# Systematic behavior of the fusion barrier parameters using the double folding model

## 1.1 Introduction

The interaction between two nuclei is governed by two potentials, the first is the repulsive Coulomb potential and the second is the attractive nuclear potential. The combination between these potentials forms the potential barrier; this barrier has to be penetrated for fusion to occur. The nucleus-nucleus potential [1- 3] plays an important role in the description of fusion in any model [3-7]. Coulomb interaction is well known from the classical treatment of the electrostatic force between charged bodies, but the nuclear contribution of the interaction potential is less known. For many studies of nucleon and light ion scattering, the major part of the nuclear interaction potential can be approximated by a Woods-Saxon (WS) form [8-11] which gives a simple analytic expression. The WS real potential combined with an imaginary part of the same radial shape, or slightly modified shape, forms the optical model potential [12-16]. This potential has been used successfully for the scattering of light ions.

Historically, the basis of the optical model was developed by comparing the results of the scattering of neutrons by nuclei to those obtained in optics for the scattering of light by transparent spheres. The first optical potentials were built for the interaction of neutrons with nuclei and afterwards for the scattering of protons [17, 18], α- particles [19] and heavy ions [5, 20-22]. The optical potential consists of two parts; the first part is a real part and it deals with the refraction, the second is an imaginary part and it deals with the absorption into reaction channels.

The interaction between heavy ions (HI) may be quite complicated; however, if we are only interested in the averaged properties, it is possible to simplify the situation by a large extent. An optical model potential for interaction between target and projectile can represent the average interaction between the incident nucleons in the projectile nucleus and nucleons in the target nucleus. It, therefore, replaces the complicated many body problem posed by the interaction of two nuclei by the much simpler problem of two particles interacting through a potential. A microscopic model of the potential may be constructed by folding the fundamental nucleon-nucleon interaction with the nuclear densities [5]. Such a folding model has been known to be quite successful in describing nucleus-nucleus interaction [5, 23, 24] data if an appropriate nucleon-nucleon interaction is used as the starting point.

Recently, the double folding model (DFM) plays an important role in the description of nuclear reactions. The DFM, which starts from realistic nuclear densities, has become one of the most popular methods for calculating the real part of the optical potential. On the basis of the DFM, detailed fits to elastic-scattering data for many systems were obtained [5, 23, 25-28], and helped in developing phenomenological potentials [23,27,31,32] to give good agreement with data.

The different nuclear density distributions can be introduced in the folding calculation. The average nuclear matter density is somewhat smaller than the density at the center of the nucleus ($\rho_0$). This is attributed to a large diffused surface region where the density drops off to zero more or less exponentially. The nuclear densities can be obtained for example from Hartree-Fock Boglioubov calculations [33, 34] or from a Quasiparticle Random Phase Approximation with Skyrme's forces [33, 35]. For many purposes, the radial distribution of nuclear density may be represented by two-parameter Fermi (2pF) distribution or three-parameter Fermi (3pF) distribution. Information about the nuclear density may be obtained from electron scattering measurements. These give directly the charge distribution from which proton distribution may be obtained. Parameters for different density distributions are calculated for the most of nuclei. In the present thesis, the tabulated parameters in reference [36] are used for the proton density with 2pF and 3pF distributions, they are given in table (1.1). Total nucleon distribution is approximated to have the same radial distribution as the proton distribution with magnitude ratio of *A/Z*.

Nucleus-nucleus potential is a function of center of mass separation distance ($R$) between the two interacting nuclei. At large separation, where the hadronic forces have become negligible, the heavy ion (HI) potential between two spherical nuclei become pure Coulomb interaction, which usually assumed to be clear and equal to

$$U_c(R) = \frac{k_e\ Z_1 e \times\ Z_2 e}{R}$$

The $R^{-1}$ dependence of $U_c(R)$ is no longer valid when the two nuclear surfaces begin to overlap at smaller values of $R$. Frequently, the Coulomb potential at small $R$ is represented by the potential felt by a point charge incident upon a charge distribution. The correct way to calculate the HI Coulomb potential is through the DFM; folding the charge density distributions with the proton-proton Coulomb potential $v_c(r_{12})$ can efficiently used to calculate $U_c(R)$ for any separation distance $R$ [5],

$$v_c(r_{12}) = \frac{k_e e^2}{r_{12}},$$

Thus the Coulomb potential between two nuclei can be calculated efficiently by double folding model [37], but the calculation of nuclear potential is more difficult to be done [5, 21, 22, 38] due to the lack of knowledge of the effective nucleon-nucleon hadronic interaction, and the mathematical complications of the many body problem. Recently many trials have been made to simplify the effective nucleon-nucleon interaction; but that which became known as M3Y [39] is probably the most widely used and certainly is representative of realistic interactions. Two versions of M3Y interaction namely, M3Y-Reid [40] and M3Y-Paris [41] effective interactions were later developed. The effective interaction depends on energy, momentum, spin, isospin and nucleons density distribution [42, 43]. Therefore, M3Y nucleon-nucleon force may have many approximate shapes as these effects are involved or not [23]. An exchange part (zero-range or finite-range) may be added to M3Y interactions [44- 46]. These M3Y interactions are purely real, so that the imaginary part of the optical potential either has to be constructed independently or, most frequently, treated phenomenologically [47].

Fusion barrier is a very important quantity in the field of super heavy elements (SHE) study. The values of barrier parameters are needed to choose the optimum conditions for SHE synthesis, and show how it is possible to be stable or decay [48-50]. Calculation of fusion-barrier is a main milestone of the present chapter. Fusion-barrier appears in the net interaction potential between two nuclei, which is the sum of all interactions. Nuclear potential has too short range compared to Coulomb potential so; the net interaction potential is positive (repulsive) at large values of ($R$). As the two nuclei become closer the total potential increases till reaches its maximum value ($V_B$) at distance ($R_B$), then it decreases rapidly to negative values when the interacting pair fuses forming a bound system of nucleons. Height of the barrier is expected to be proportional to $Z_1Z_2$, which stands from the classical definition of coulomb potential [51]. The behavior of barrier height for different interacting pairs may deviate from the well defined proportionality relation, mentioned above, for one or more reason, which will be discussed within the present chapter. The radius of the potential barrier is also expected to increase as the sum of interacting pair radii ($R_1$+$R_2$).

Detailed study of the potential barrier parameters will be performed in the present chapter to understand their dependence on the entrance channel of the fusion reaction.

The aim of the present chapter is to study the systematic behavior of the fusion barrier parameters for large number of interacting ion pairs and its dependence on the composition of the interacting nuclei looking for simple and direct analytical expression for calculation of the interaction barrier starting from masses, charges, and radii of interacting nuclei. DFM will be used to drive the interaction potential, starting from empirical nuclear density and M3Y nucleon-nucleon force, which will be represented briefly in the next sections.

## 1.2 Double folding model (DFM)

It is generally assumed that the interaction $U(\boldsymbol{R})$ is a sum of local two-body potential $v(r_{12})$, although many-body aspects may be represented by a dependence of $v(r_{12})$ on the density of the nuclear matter in which the two interacting nucleons are embedded [5]. Then the folded potential may be written as

$$U(\boldsymbol{R}) = \int d\boldsymbol{r_1} \int d\boldsymbol{r_2}\ \rho_1(\boldsymbol{r_1}) v(\boldsymbol{r_{12}}) \rho_2(\boldsymbol{r_2}) \qquad (1.1)$$

$$\boldsymbol{r_{12}} = |\boldsymbol{R} + \boldsymbol{r_2} - \boldsymbol{r_1}|$$

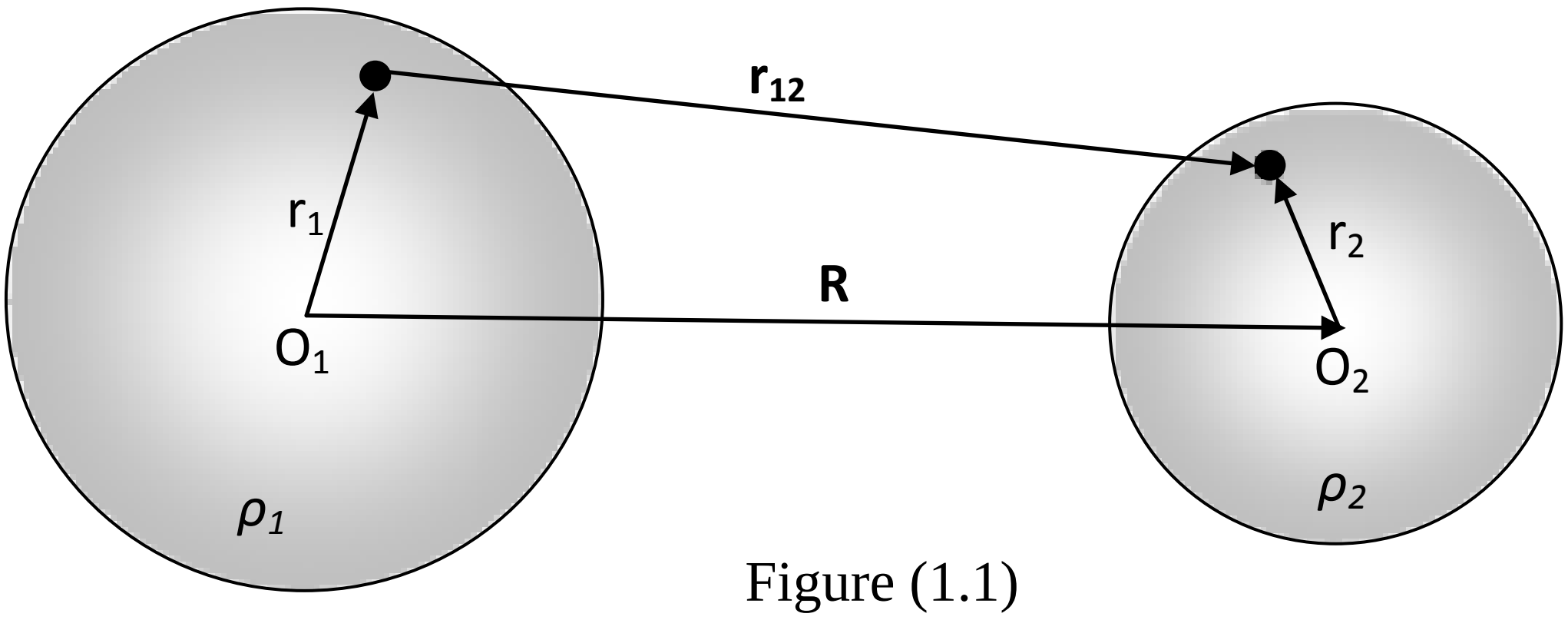

Figure (1.1)

The coordinates are defined in Figure (1.1), where $R$ is the distance between the mass centers of the two interacting nuclei. Here $\rho_1(\boldsymbol{r_1})$ and $\rho_2(\boldsymbol{r_2})$ are the density distributions of the projectile and target ground states respectively, normalized so that

$$\int \rho_i(\boldsymbol{r})d\boldsymbol{r} = X_i \qquad ,(i = 1,2) \tag{1.2}$$

where $X$ is the number of nucleons which are sensitive for the interaction $v(r_{12})$, i.e. $X$ is the number of protons for Coulomb interaction, and the mass number for nuclear interaction. For a scalar potential $v(r_{12})$ and if density distributions of both nuclei are taken to be spherically symmetric, then the folded potential $U(\boldsymbol{R})$ is spherically symmetric; if one or both densities are nonspherical, $U(\boldsymbol{R})$ will be nonspherical.

The calculation of nucleus-nucleus interaction using the double folding model, given by equation (1.1), includes complicated integrations and is not easy to evaluate, but it becomes easy to calculate if we work in momentum space [5,52]; the double folding reduces to a product of three Fourier transforms; see (appendix 1-A**)**. Often the Fourier transform of the effective nucleon-nucleon potential $v(r_{12})$ has an analytic form and need not to done numerically; see (appendix 1-B**).**

From the convolution theorem the Fourier transform of the folded quantity is simply the product of the transforms of the individual component functions. This makes the calculation much easier than directly doing the folding integrals

$$\widetilde{U}(\boldsymbol{k}) = \tilde{\rho}_1(\boldsymbol{k})\tilde{v}(\boldsymbol{k})\tilde{\rho}_2(-\boldsymbol{k}) \qquad (1.3)$$

If the density distribution is spherically symmetric, we can apply the simplified expression of Fourier transform by integrating over the solid angle to get,

$$\tilde{\rho}(k) = 4\pi \int j_0(kr)\, \rho(r)\, r^2\, dr \qquad (1.4)$$

So that

$$\widetilde{U}(\boldsymbol{k}) = \left[4\pi \int j_0(kr_1)\, \rho_1(r_1)\, r_1^2\, dr_1\right] \tilde{v}(\boldsymbol{k}) \left[4\pi \int j_0(kr_2)\, \rho_2(r_2) r_2^2\, dr_2\right] \quad (\mathbf{1.5})$$

The potential $\tilde{v}(k)$ has an analytic form, and each of the two square brackets in equation (1.5) can be calculated separately, then there is no cross terms. The folded potential $U(\boldsymbol{R})$ is the back Fourier transformation of the total potential $\widetilde{U}(\boldsymbol{k})$.

$$U(\boldsymbol{R}) = (2\pi)^{-3} \int \widetilde{U}(\boldsymbol{k})\, e^{(-i\boldsymbol{k}.\boldsymbol{R})}\, d\boldsymbol{k}.$$

$$U(R) = 8 \int k^2\, j_0(kR)\, \tilde{v}(k) \left[\int j_0(kr_1) \rho_1(r_1) r_1^2\, dr_1\right] \left[\int j_0(kr_2)\, \rho_2(r_2) r_2^2 dr_2\right] dk \quad (1.6)$$

This integral is easy to compute numerically, because it consists of three integrals each in one dimension.

## 1.3 Details of calculation

For all interactions the units of MeV and *fm* are used for the strengths of the interactions and the lengths respectively. In this work the nuclear interaction used is the well-known M3Y-Reid force [5, 37] in the form

$$v_n(r) = \frac{7999\, e^{-4r}}{4r} - \frac{2134\, e^{-2.5r}}{2.5r} - 262\delta(r) \qquad MeV \qquad (1.7)$$

making use of the Fourier transform; see (appendix 1-B.2**)**

$$\tilde{v}_n(k) = \frac{7999 \times 4\pi}{4[k^2 + 4^2]} - \frac{2134 \times 4\pi}{2.5[k^2 + 2.5^2]} - 262 \qquad MeV\, fm^3 \quad (1.8)$$

Coulomb interaction between two charged nucleons (protons) in the two nuclei separated by distance ($r$) is given by

$$v_c(r) = (1.44)\left(\frac{1}{r}\right) \qquad MeV \qquad (1.9)$$

and its Fourier transform; see (appendix 1-B.1**)**

$$\tilde{v}_c(k) = \frac{4\pi \times 1.44}{k^2} \qquad MeV\, fm^3 \qquad (1.10)$$

The charge and matter density distribution can be described in the form of 2pF or 3pF distributions, given respectively by

$$\rho(r) = \frac{\rho_o}{1 + e^{\frac{r - R_o}{a}}} \qquad (1.11 - A)$$

$$\rho(r) = \frac{\rho_o\left[1 + w\left(\frac{r}{R_0}\right)^2\right]}{1 + e^{\frac{r - R_o}{a}}} \qquad (1.11 - B)$$

where $a$ is a parameter that measures the "diffuseness" of the nuclear surface, with typical values around 0.5 *fm*, and $R_0$ is the nuclear radius. The 2pF is the same as 3pF with ($w$ = 0). Density distributions parameters are used as obtained from elastic electron scattering experiments [36] and are presented in table (1.1).

The net potential between two interacting nuclei is a sum of two parts, repulsive part plus attractive part.

$$U_{Net}(R) = U_{Coulomb}(R) + U_{Nuclear}(R)$$

each part in the above equation is a nucleus-nucleus interaction calculated in the frame work of the double folding model given by equation (1.6), then added to obtain the total interaction potential between the interacting pair. The radial distribution of the potential depends on the composition, shape, and orientation of the two interacting nuclei [53]. Similarly, the potential barrier parameters ($R_B$ and $V_B$) depend on the same variables. The aim of this work is to study the dependence of Coulomb barrier parameters on the compositions of the target nucleus and the projectile nucleus, assuming that both nuclei have spherical shape and have ground state density described by 2pF or 3pF distributions. On the basis of the presented treatment, behavior of potential barrier parameters will be discussed to show their systematic behavior with the composition of the interacting pair.

## 1.4 Numerical calculations and results

Table (1.1) density-distribution parameters obtained from elastic electron scattering [36]

| Nucleus | $R_0$ fm | a fm | w | Nucleus | $R_0$ fm | a fm | w |
|---|---|---|---|---|---|---|---|
| $^{16}O$ | 2.608 | 0.513 | -0.051 | $^{68}Zn$ | 4.353 | 0.567 | |
| $^{19}F$ | 2.59 | 0.564 | | $^{70}Zn$ | 4.409 | 0.583 | |
| $^{20}Ne$ | 2.74 | 0.569 | | $^{88}Sr$ | 4.83 | 0.496 | |
| $^{22}Ne$ | 2.782 | 0.549 | | $^{89}Y$ | 4.86 | 0.542 | |
| $^{24}Mg$ | 3.192 | 0.604 | -0.249 | $^{93}Nb$ | 4.87 | 0.573 | |
| $^{25}Mg$ | 2.76 | 0.608 | | $^{110}Cd$ | 5.33 | 0.535 | |
| $^{26}Mg$ | 3.05 | 0.524 | | $^{112}Cd$ | 5.38 | 0.532 | |
| $^{27}Al$ | 2.84 | 0.569 | | $^{114}Cd$ | 5.40 | 0.537 | |
| $^{28}Si$ | 3.30 | 0.545 | -0.18 | $^{116}Cd$ | 5.42 | 0.532 | |
| $^{29}Si$ | 3.17 | 0.52 | | In | 5.24 | 0.52 | |
| $^{31}P$ | 3.353 | 0.5789 | -0.160 | 112Sn | 5.375 | 0.56 | |
| $^{32}S$ | 3.458 | 0.6098 | -0.208 | $^{116}Sn$ | 5.416 | 0.552 | |
| $^{40}Ar$ | 3.73 | 0.62 | -0.19 | $^{118}Sn$ | 5.442 | 0.543 | |
| $^{39}K$ | 3.743 | 0.585 | -0.201 | $^{120}Sn$ | 5.32 | 0.576 | |
| $^{40}Ca$ | 3.766 | 0.586 | -0.161 | $^{124}Sn$ | 5.490 | 0.534 | |
| Ti | 3.75 | 0.567 | | Sb | 5.32 | 0.57 | |
| $^{51}V$ | 3.91 | 0.532 | | La | 5.71 | 0.535 | |
| Cr | 3.975 | 0.53 | | $^{142}Nd$ | 5.6135 | 0.5868 | 0.096 |
| $^{55}Mn$ | 3.89 | 0.567 | | $^{144}Nd$ | 5.6256 | 0.6178 | |
| Fe | 3.98 | 0.569 | | $^{146}Nd$ | 5.867 | 0.556 | |
| $^{54}Fe$ | 4.012 | 0.5339 | | $^{148}Nd$ | 5.6703 | 0.644 | |
| $^{56}Fe$ | 3.971 | 0.5935 | | $^{150}Nd$ | 5.865 | 0.571 | |
| $^{58}Fe$ | 4.027 | 0.5757 | | $^{148}Sm$ | 5.771 | 0.596 | |
| $^{59}Co$ | 4.08 | 0.569 | | $^{154}Sm$ | 5.9387 | 0.522 | |
| Ni | 4.09 | 0.569 | | $^{165}Ho$ | 6.12 | 0.57 | |
| $^{58}Ni$ | 4.3092 | 0.5169 | | $^{181}Ta$ | 6.38 | 0.64 | |
| $^{60}Ni$ | 4.4891 | 0.5369 | | $^{184}W$ | 6.51 | 0.535 | |
| $^{61}Ni$ | 4.4024 | 0.5401 | | $^{186}W$ | 6.58 | 0.480 | |
| $^{62}Ni$ | 4.4425 | 0.5386 | | $^{197}Au$ | 6.38 | 0.535 | |
| $^{64}Ni$ | 4.5211 | 0.5278 | | P[]b | 6.69 | 0.494 | |
| Cu | 4.2 | 0.569 | | $^{206}Pb$ | 6.61 | 0.545 | |
| $^{63}Cu$ | 4.214 | 0.586 | | $^{207}Pb$ | 6.62 | 0.546 | |
| $^{65}Cu$ | 4.271 | 0.579 | | $^{208}Pb$ | 6.624 | 0.549 | |
| Zn | 4.28 | 0.569 | | $^{209}Bi$ | 6.75 | 0.468 | |
| $^{64}Zn$ | 4.285 | 0.584 | | $^{232}Th$ | 6.7915 | 0.571 | |
| $^{66}Zn$ | 4.286 | 0.595 | | $^{238}U$ | 6.854 | 0.605 | |

In table (1.1), the parameters are tabulated for 2pF and 3pF distributions. The parameters tabulated without mass number are the results for targets of natural isotopic composition.

For trans-uranium elements ($Z \geq 93$) the density distribution parameters are approximated as following:

1) The radius parameter ($R_0$) [6]

$$R_0 = 1.28A^{1/3} - 0.76 + 0.8\,A^{-1/3} \qquad fm$$

2) The diffuseness parameter is selected to be (0.54 *fm*) for all trans-uranium elements

Density distribution for $^4He$ projectile [5] is taken as

$$Matter\ density = 0.4229\,e^{-0.7024\,r^2} \qquad fm^{-3}$$

$$Charge\ density = 0.5\ \times Matter\ density \qquad fm^{-3}$$

It is given in tables (1.2- *a, b, c, d*) the results for the barrier radius ($R_B$) and the barrier height ($V_B$) calculated by using DFM, and the values of Coulomb and nuclear interactions at separation $R_B$, donated respectively by $V_C$ and $V_N$, followed by graphs and fits showing the systematic behavior of the fusion barrier parameters.

Table (1.2-*a*) Barrier position ($R_B$), barrier height ($V_B$), Coulomb interaction at $R$=$R_B$ ($V_C$,), and nuclear interaction at $R$=$R_B$ ($V_N$), for the reactions between different targets and $^4He$ as projectile.

| $Z_T$ | $A_T$ | $R_B$ fm | $V_B$ MeV | $V_C$ MeV | $V_N$ MeV | $Z_T$ | $A_T$ | $R_B$ fm | $V_B$ MeV | $V_C$ MeV | $V_N$ MeV |
|---|---|---|---|---|---|---|---|---|---|---|---|
| **8** | **16** | 7.313 | 2.907 | 3.149 | -0.242 | **50** | **118** | 9.638 | 13.966 | 14.930 | -0.963 |
| **9** | **19** | 7.638 | 3.122 | 3.392 | -0.270 | **50** | **120** | 9.663 | 13.898 | 14.891 | -0.993 |
| **10** | **20** | 7.688 | 3.439 | 3.745 | -0.306 | **50** | **124** | 9.688 | 13.915 | 14.852 | -0.937 |
| **10** | **22** | 7.713 | 3.442 | 3.733 | -0.291 | **51** | **121** | 9.638 | 14.223 | 15.229 | -1.007 |
| **12** | **24** | 7.563 | 4.302 | 4.564 | -0.262 | **57** | **139** | 9.863 | 15.605 | 16.630 | -1.025 |
| **12** | **25** | 7.888 | 4.010 | 4.379 | -0.369 | **60** | **142** | 9.988 | 16.146 | 17.290 | -1.144 |
| **12** | **26** | 7.763 | 4.113 | 4.450 | -0.337 | **60** | **144** | 10.038 | 16.030 | 17.200 | -1.169 |
| **13** | **27** | 7.788 | 4.424 | 4.805 | -0.381 | **60** | **146** | 10.038 | 16.110 | 17.204 | -1.094 |
| **14** | **28** | 7.688 | 4.874 | 5.240 | -0.366 | **60** | **148** | 10.188 | 15.781 | 16.946 | -1.164 |
| **14** | **29** | 7.788 | 4.782 | 5.175 | -0.393 | **60** | **150** | 10.113 | 15.993 | 17.076 | -1.083 |
| **15** | **31** | 7.888 | 5.086 | 5.472 | -0.386 | **62** | **148** | 10.063 | 16.539 | 17.734 | -1.195 |
| **16** | **32** | 7.913 | 5.420 | 5.815 | -0.395 | **62** | **154** | 10.013 | 16.738 | 17.821 | -1.083 |
| **18** | **40** | 8.288 | 5.825 | 6.249 | -0.425 | **67** | **165** | 10.288 | 17.544 | 18.742 | -1.198 |
| **19** | **39** | 8.113 | 6.271 | 6.740 | -0.469 | **73** | **181** | 10.738 | 18.282 | 19.563 | -1.281 |
| **20** | **40** | 8.188 | 6.524 | 7.031 | -0.507 | **74** | **184** | 10.513 | 19.036 | 20.255 | -1.219 |
| **22** | **48** | 8.388 | 6.978 | 7.550 | -0.573 | **74** | **186** | 10.438 | 19.263 | 20.401 | -1.139 |
| **23** | **51** | 8.388 | 7.322 | 7.892 | -0.570 | **79** | **197** | 10.413 | 20.519 | 21.833 | -1.313 |
| **24** | **52** | 8.413 | 7.622 | 8.211 | -0.589 | **82** | **206** | 10.638 | 20.852 | 22.180 | -1.327 |
| **25** | **55** | 8.513 | 7.838 | 8.453 | -0.614 | **82** | **207** | 10.663 | 20.824 | 22.128 | -1.304 |
| **26** | **54** | 8.438 | 8.243 | 8.869 | -0.625 | **82** | **208** | 10.663 | 20.790 | 22.128 | -1.337 |
| **26** | **56** | 8.638 | 8.009 | 8.663 | -0.654 | **83** | **209** | 10.538 | 21.398 | 22.664 | -1.265 |
| **26** | **58** | 8.638 | 8.024 | 8.664 | -0.640 | **90** | **232** | 10.888 | 22.356 | 23.787 | -1.431 |
| **27** | **59** | 8.638 | 8.339 | 8.997 | -0.658 | **92** | **238** | 11.038 | 22.487 | 23.982 | -1.495 |
| **28** | **58** | 8.463 | 8.867 | 9.523 | -0.656 | **93** | **239** | 11.238 | 22.467 | 23.811 | -1.343 |
| **28** | **60** | 8.538 | 8.841 | 9.437 | -0.596 | **94** | **239** | 11.238 | 22.723 | 24.067 | -1.343 |
| **28** | **61** | 8.563 | 8.778 | 9.411 | -0.633 | **95** | **243** | 11.263 | 22.885 | 24.268 | -1.384 |
| **28** | **62** | 8.588 | 8.757 | 9.383 | -0.626 | **96** | **245** | 11.288 | 23.098 | 24.475 | -1.377 |
| **28** | **64** | 8.638 | 8.736 | 9.329 | -0.593 | **97** | **247** | 11.313 | 23.305 | 24.674 | -1.369 |
| **29** | **63** | 8.788 | 8.797 | 9.498 | -0.701 | **98** | **249** | 11.313 | 23.512 | 24.928 | -1.416 |
| **29** | **65** | 8.838 | 8.766 | 9.444 | -0.678 | **99** | **254** | 11.363 | 23.650 | 25.073 | -1.423 |
| **30** | **64** | 8.813 | 9.073 | 9.797 | -0.724 | **100** | **253** | 11.363 | 23.925 | 25.324 | -1.400 |
| **30** | **66** | 8.888 | 9.003 | 9.715 | -0.711 | **101** | **255** | 11.363 | 24.130 | 25.577 | -1.447 |
| **30** | **68** | 8.863 | 9.055 | 9.742 | -0.687 | **102** | **255** | 11.363 | 24.384 | 25.830 | -1.447 |
| **30** | **70** | 8.988 | 8.928 | 9.606 | -0.677 | **103** | **257** | 11.388 | 24.589 | 26.027 | -1.438 |
| **38** | **88** | 8.988 | 11.375 | 12.169 | -0.795 | **104** | **261** | 11.413 | 24.746 | 26.222 | -1.476 |
| **39** | **89** | 9.163 | 11.435 | 12.251 | -0.815 | **105** | **262** | 11.413 | 24.974 | 26.474 | -1.500 |
| **41** | **93** | 9.263 | 11.859 | 12.739 | -0.880 | **106** | **263** | 11.438 | 25.204 | 26.669 | -1.466 |
| **48** | **110** | 9.513 | 13.596 | 14.522 | -0.926 | **107** | **262** | 11.413 | 25.478 | 26.978 | -1.500 |
| **48** | **112** | 9.538 | 13.542 | 14.484 | -0.942 | **108** | **265** | 11.438 | 25.658 | 27.172 | -1.514 |
| **48** | **114** | 9.588 | 13.477 | 14.407 | -0.930 | **109** | **266** | 11.438 | 25.883 | 27.421 | -1.538 |
| **48** | **116** | 9.613 | 13.466 | 14.370 | -0.904 | **110** | **269** | 11.463 | 26.062 | 27.614 | -1.552 |
| **49** | **115** | 9.388 | 14.038 | 15.022 | -0.984 | **111** | **272** | 11.488 | 26.238 | 27.803 | -1.565 |
| **50** | **112** | 9.588 | 13.997 | 15.008 | -1.011 | **112** | **277** | 11.538 | 26.366 | 27.931 | -1.565 |
| **50** | **116** | 9.638 | 13.962 | 14.929 | -0.968 | **114** | **289** | 11.663 | 26.578 | 28.126 | -1.548 |

Table (1.2-*b*) The same as Table (1.2-*a*) but for $^{16}O$ projectile.

| $Z_T$ | $A_T$ | $R_B$ fm | $V_B$ MeV | $V_C$ MeV | $V_N$ MeV | $Z_T$ | $A_T$ | $R_B$ fm | $V_B$ MeV | $V_C$ MeV | $V_N$ MeV |
|---|---|---|---|---|---|---|---|---|---|---|---|
| **8** | **16** | 8.488 | 10.006 | 10.849 | -0.842 | **50** | **118** | 10.638 | 50.309 | 54.089 | -3.781 |
| **9** | **19** | 8.738 | 10.902 | 11.855 | -0.953 | **50** | **120** | 10.663 | 50.194 | 53.963 | -3.769 |
| **10** | **20** | 8.788 | 12.044 | 13.099 | -1.055 | **50** | **124** | 10.713 | 50.083 | 53.710 | -3.627 |
| **10** | **22** | 8.838 | 12.012 | 13.025 | -1.013 | **51** | **121** | 10.638 | 51.333 | 55.175 | -3.843 |
| **12** | **24** | 8.713 | 14.748 | 15.838 | -1.090 | **57** | **139** | 10.863 | 56.309 | 60.380 | -4.071 |
| **12** | **25** | 8.963 | 14.154 | 15.412 | -1.258 | **60** | **142** | 10.938 | 58.649 | 63.137 | -4.489 |
| **12** | **26** | 8.888 | 14.338 | 15.541 | -1.203 | **60** | **144** | 10.988 | 58.347 | 62.832 | -4.485 |
| **13** | **27** | 8.863 | 15.515 | 16.884 | -1.370 | **60** | **146** | 11.038 | 58.362 | 62.565 | -4.203 |
| **14** | **28** | 8.813 | 16.891 | 18.279 | -1.388 | **60** | **148** | 11.113 | 57.625 | 62.124 | -4.499 |
| **14** | **29** | 8.913 | 16.688 | 18.080 | -1.392 | **60** | **150** | 11.088 | 58.024 | 62.280 | -4.255 |
| **15** | **31** | 8.988 | 17.756 | 19.204 | -1.449 | **62** | **148** | 11.038 | 60.135 | 64.652 | -4.517 |
| **16** | **32** | 9.013 | 18.913 | 20.415 | -1.502 | **62** | **154** | 11.013 | 60.443 | 64.792 | -4.349 |
| **18** | **40** | 9.363 | 20.514 | 22.120 | -1.607 | **67** | **165** | 11.263 | 63.803 | 68.458 | -4.655 |
| **19** | **39** | 9.213 | 21.997 | 23.732 | -1.736 | **73** | **181** | 11.638 | 67.110 | 72.179 | -5.069 |
| **20** | **40** | 9.263 | 22.986 | 24.851 | -1.865 | **74** | **184** | 11.513 | 69.223 | 73.962 | -4.739 |
| **22** | **48** | 9.463 | 24.738 | 26.762 | -2.023 | **74** | **186** | 11.438 | 69.702 | 74.449 | -4.748 |
| **23** | **51** | 9.463 | 25.865 | 27.975 | -2.110 | **79** | **197** | 11.388 | 74.525 | 79.830 | -5.306 |
| **24** | **52** | 9.488 | 26.942 | 29.115 | -2.173 | **82** | **206** | 11.613 | 75.969 | 81.245 | -5.276 |
| **25** | **55** | 9.538 | 27.837 | 30.168 | -2.331 | **82** | **207** | 11.638 | 75.877 | 81.072 | -5.195 |
| **26** | **54** | 9.488 | 29.175 | 31.539 | -2.363 | **82** | **208** | 11.638 | 75.778 | 81.071 | -5.292 |
| **26** | **56** | 9.663 | 28.588 | 30.968 | -2.380 | **83** | **209** | 11.563 | 77.469 | 82.594 | -5.125 |
| **26** | **58** | 9.688 | 28.578 | 30.890 | -2.312 | **90** | **232** | 11.838 | 81.773 | 87.486 | -5.713 |
| **27** | **59** | 9.663 | 29.693 | 32.161 | -2.468 | **92** | **238** | 11.963 | 82.560 | 88.486 | -5.926 |
| **28** | **58** | 9.538 | 31.299 | 33.789 | -2.490 | **93** | **239** | 12.188 | 82.300 | 87.794 | -5.493 |
| **28** | **60** | 9.613 | 31.150 | 33.516 | -2.366 | **94** | **239** | 12.188 | 83.244 | 88.738 | -5.493 |
| **28** | **61** | 9.638 | 31.013 | 33.435 | -2.422 | **95** | **243** | 12.213 | 83.862 | 89.496 | -5.634 |
| **28** | **62** | 9.663 | 30.933 | 33.347 | -2.414 | **96** | **245** | 12.238 | 84.659 | 90.272 | -5.613 |
| **28** | **64** | 9.713 | 30.825 | 33.175 | -2.350 | **97** | **247** | 12.238 | 85.435 | 91.211 | -5.776 |
| **29** | **63** | 9.813 | 31.450 | 34.012 | -2.562 | **98** | **249** | 12.263 | 86.210 | 91.962 | -5.751 |
| **29** | **65** | 9.863 | 31.327 | 33.840 | -2.513 | **99** | **254** | 12.313 | 86.746 | 92.526 | -5.780 |
| **30** | **64** | 9.838 | 32.454 | 35.094 | -2.640 | **100** | **253** | 12.288 | 87.756 | 93.645 | -5.889 |
| **30** | **66** | 9.888 | 32.254 | 34.919 | -2.664 | **101** | **255** | 12.313 | 88.527 | 94.388 | -5.861 |
| **30** | **68** | 9.888 | 32.330 | 34.917 | -2.587 | **102** | **255** | 12.313 | 89.461 | 95.323 | -5.861 |
| **30** | **70** | 9.988 | 31.970 | 34.567 | -2.597 | **103** | **257** | 12.313 | 90.232 | 96.258 | -6.026 |
| **38** | **88** | 10.063 | 40.442 | 43.464 | -3.021 | **104** | **261** | 12.363 | 90.834 | 96.800 | -5.966 |
| **39** | **89** | 10.188 | 40.929 | 44.060 | -3.131 | **105** | **262** | 12.363 | 91.682 | 97.729 | -6.047 |
| **41** | **93** | 10.263 | 42.625 | 45.978 | -3.354 | **106** | **263** | 12.363 | 92.537 | 98.668 | -6.131 |
| **48** | **110** | 10.513 | 48.877 | 52.546 | -3.669 | **107** | **262** | 12.338 | 93.545 | 99.792 | -6.247 |
| **48** | **112** | 10.563 | 48.686 | 52.298 | -3.612 | **108** | **265** | 12.363 | 94.228 | 100.527 | -6.299 |
| **48** | **114** | 10.613 | 48.489 | 52.049 | -3.560 | **109** | **266** | 12.363 | 95.065 | 101.448 | -6.383 |
| **48** | **116** | 10.613 | 48.426 | 52.048 | -3.623 | **110** | **269** | 12.388 | 95.745 | 102.178 | -6.433 |
| **49** | **115** | 10.438 | 50.315 | 54.027 | -3.712 | **111** | **272** | 12.413 | 96.413 | 102.895 | -6.481 |
| **50** | **112** | 10.588 | 50.500 | 54.344 | -3.845 | **112** | **277** | 12.463 | 96.914 | 103.401 | -6.488 |
| **50** | **116** | 10.638 | 50.337 | 54.088 | -3.751 | **114** | **289** | 12.588 | 97.764 | 104.206 | -6.441 |

Table (*1.2-c*) The same as Table (1.2-*a*) but for $^{40}Ca$ projectile.

| $Z_T$ | $A_T$ | $R_B$ fm | $V_B$ MeV | $V_C$ MeV | $V_N$ MeV | $Z_T$ | $A_T$ | $R_B$ fm | $V_B$ MeV | $V_C$ MeV | $V_N$ MeV |
|---|---|---|---|---|---|---|---|---|---|---|---|
| **8** | **16** | 9.263 | 22.986 | 24.851 | -1.865 | **50** | **118** | 11.338 | 117.998 | 126.866 | -8.868 |
| **9** | **19** | 9.488 | 25.170 | 27.294 | -2.124 | **50** | **120** | 11.338 | 117.827 | 126.867 | -9.040 |
| **10** | **20** | 9.513 | 27.838 | 30.251 | -2.414 | **50** | **124** | 11.413 | 117.454 | 126.031 | -8.577 |
| **10** | **22** | 9.588 | 27.740 | 30.015 | -2.274 | **51** | **121** | 11.313 | 120.468 | 129.699 | -9.231 |
| **12** | **24** | 9.488 | 33.857 | 36.358 | -2.501 | **57** | **139** | 11.563 | 132.216 | 141.802 | -9.585 |
| **12** | **25** | 9.688 | 32.811 | 35.645 | -2.834 | **60** | **142** | 11.613 | 138.037 | 148.659 | -10.621 |
| **12** | **26** | 9.638 | 33.108 | 35.827 | -2.719 | **60** | **144** | 11.663 | 137.440 | 147.980 | -10.540 |
| **13** | **27** | 9.613 | 35.884 | 38.916 | -3.032 | **60** | **146** | 11.713 | 137.272 | 147.388 | -10.116 |
| **14** | **28** | 9.588 | 38.929 | 42.001 | -3.072 | **60** | **148** | 11.788 | 135.927 | 146.407 | -10.480 |
| **14** | **29** | 9.663 | 38.552 | 41.690 | -3.138 | **60** | **150** | 11.763 | 136.561 | 146.755 | -10.195 |
| **15** | **31** | 9.738 | 41.030 | 44.310 | -3.280 | **62** | **148** | 11.713 | 141.616 | 152.304 | -10.689 |
| **16** | **32** | 9.763 | 43.696 | 47.111 | -3.416 | **62** | **154** | 11.713 | 142.020 | 152.290 | -10.270 |
| **18** | **40** | 10.088 | 47.584 | 51.322 | -3.738 | **67** | **165** | 11.938 | 150.345 | 161.457 | -11.111 |
| **19** | **39** | 9.938 | 50.941 | 54.998 | -4.057 | **73** | **181** | 12.288 | 158.785 | 170.893 | -12.108 |
| **20** | **40** | 9.988 | 53.316 | 57.614 | -4.298 | **74** | **184** | 12.188 | 163.207 | 174.655 | -11.448 |
| **22** | **48** | 10.188 | 57.526 | 62.138 | -4.613 | **74** | **186** | 12.138 | 164.059 | 175.379 | -11.320 |
| **23** | **51** | 10.213 | 60.087 | 64.798 | -4.711 | **79** | **197** | 12.088 | 175.593 | 188.008 | -12.415 |
| **24** | **52** | 10.213 | 62.604 | 67.616 | -5.012 | **82** | **206** | 12.288 | 179.263 | 191.944 | -12.681 |
| **25** | **55** | 10.263 | 64.787 | 70.088 | -5.300 | **82** | **207** | 12.313 | 179.060 | 191.556 | -12.496 |
| **26** | **54** | 10.213 | 67.820 | 73.244 | -5.424 | **82** | **208** | 12.313 | 178.850 | 191.554 | -12.704 |
| **26** | **56** | 10.363 | 66.664 | 72.185 | -5.521 | **83** | **209** | 12.263 | 182.423 | 194.685 | -12.263 |
| **26** | **58** | 10.388 | 66.599 | 72.015 | -5.416 | **90** | **232** | 12.513 | 193.304 | 206.904 | -13.600 |
| **27** | **59** | 10.388 | 69.191 | 74.785 | -5.594 | **92** | **238** | 12.638 | 195.470 | 209.387 | -13.916 |
| **28** | **58** | 10.288 | 72.720 | 78.310 | -5.590 | **93** | **239** | 12.863 | 194.792 | 207.953 | -13.161 |
| **28** | **60** | 10.338 | 72.348 | 77.909 | -5.561 | **94** | **239** | 12.838 | 197.030 | 210.597 | -13.568 |
| **28** | **61** | 10.388 | 72.095 | 77.548 | -5.453 | **95** | **243** | 12.888 | 198.526 | 212.006 | -13.480 |
| **28** | **62** | 10.413 | 71.913 | 77.358 | -5.446 | **96** | **245** | 12.888 | 200.433 | 214.284 | -13.850 |
| **28** | **64** | 10.463 | 71.649 | 76.986 | -5.337 | **97** | **247** | 12.913 | 202.292 | 216.092 | -13.800 |
| **29** | **63** | 10.513 | 73.411 | 79.363 | -5.952 | **98** | **249** | 12.938 | 204.145 | 217.893 | -13.748 |
| **29** | **65** | 10.563 | 73.122 | 78.988 | -5.866 | **99** | **254** | 12.988 | 205.459 | 219.279 | -13.820 |
| **30** | **64** | 10.538 | 75.776 | 81.902 | -6.126 | **100** | **253** | 12.963 | 207.847 | 221.906 | -14.059 |
| **30** | **66** | 10.588 | 75.358 | 81.520 | -6.163 | **101** | **255** | 12.963 | 209.691 | 224.121 | -14.430 |
| **30** | **68** | 10.588 | 75.450 | 81.515 | -6.065 | **102** | **255** | 12.963 | 211.911 | 226.341 | -14.430 |
| **30** | **70** | 10.713 | 74.701 | 80.563 | -5.862 | **103** | **257** | 12.988 | 213.755 | 228.123 | -14.368 |
| **38** | **88** | 10.788 | 94.320 | 101.350 | -7.030 | **104** | **261** | 13.013 | 215.220 | 229.896 | -14.676 |
| **39** | **89** | 10.888 | 95.679 | 103.061 | -7.382 | **105** | **262** | 13.013 | 217.240 | 232.104 | -14.864 |
| **41** | **93** | 10.963 | 99.799 | 107.599 | -7.800 | **106** | **263** | 13.013 | 219.277 | 234.333 | -15.055 |
| **48** | **110** | 11.213 | 114.528 | 123.156 | -8.628 | **107** | **262** | 12.988 | 221.662 | 236.980 | -15.318 |
| **48** | **112** | 11.263 | 114.099 | 122.611 | -8.513 | **108** | **265** | 13.013 | 223.309 | 238.748 | -15.439 |
| **48** | **114** | 11.313 | 113.673 | 122.062 | -8.389 | **109** | **266** | 13.013 | 225.304 | 240.937 | -15.633 |
| **48** | **116** | 11.338 | 113.514 | 121.793 | -8.280 | **110** | **269** | 13.063 | 226.945 | 242.230 | -15.286 |
| **49** | **115** | 11.138 | 117.760 | 126.571 | -8.811 | **111** | **272** | 13.088 | 228.559 | 243.954 | -15.395 |
| **50** | **112** | 11.288 | 118.494 | 127.428 | -8.934 | **112** | **277** | 13.138 | 229.791 | 245.207 | -15.416 |
| **50** | **116** | 11.338 | 118.095 | 126.863 | -8.768 | **114** | **289** | 13.263 | 231.912 | 247.241 | -15.329 |

Table (*1.2-d*) The same as Table (1.2-*a*) but for $^{60}Ni$ projectile.

| $Z_T$ | $A_T$ | $R_B$ fm | $V_B$ MeV | $V_C$ MeV | $V_N$ MeV | $Z_T$ | $A_T$ | $R_B$ fm | $V_B$ MeV | $V_C$ MeV | $V_N$ MeV |
|---|---|---|---|---|---|---|---|---|---|---|---|
| **8** | **16** | 9.613 | 31.150 | 33.516 | -2.366 | **50** | **118** | 11.713 | 160.648 | 171.882 | -11.234 |
| **9** | **19** | 9.838 | 34.132 | 36.842 | -2.710 | **50** | **120** | 11.713 | 160.430 | 171.884 | -11.453 |
| **10** | **20** | 9.888 | 37.746 | 40.736 | -2.990 | **50** | **124** | 11.763 | 159.941 | 171.149 | -11.207 |
| **10** | **22** | 9.938 | 37.631 | 40.531 | -2.900 | **51** | **121** | 11.688 | 164.012 | 175.707 | -11.696 |
| **12** | **24** | 9.838 | 45.875 | 49.077 | -3.201 | **57** | **139** | 11.913 | 180.089 | 192.640 | -12.552 |
| **12** | **25** | 10.038 | 44.512 | 48.151 | -3.639 | **60** | **142** | 11.988 | 188.056 | 201.562 | -13.506 |
| **12** | **26** | 10.013 | 44.914 | 48.267 | -3.353 | **60** | **144** | 12.013 | 187.265 | 201.086 | -13.821 |
| **13** | **27** | 9.963 | 48.671 | 52.555 | -3.885 | **60** | **146** | 12.088 | 187.037 | 199.892 | -12.855 |
| **14** | **28** | 9.938 | 52.779 | 56.717 | -3.938 | **60** | **148** | 12.138 | 185.266 | 199.011 | -13.745 |
| **14** | **29** | 10.013 | 52.291 | 56.312 | -4.021 | **60** | **150** | 12.138 | 186.103 | 199.060 | -12.956 |
| **15** | **31** | 10.088 | 55.653 | 59.867 | -4.215 | **62** | **148** | 12.063 | 192.962 | 206.988 | -14.026 |
| **16** | **32** | 10.113 | 59.256 | 63.657 | -4.402 | **62** | **154** | 12.088 | 193.505 | 206.538 | -13.033 |
| **18** | **40** | 10.438 | 64.600 | 69.425 | -4.825 | **67** | **165** | 12.313 | 204.945 | 219.098 | -14.153 |
| **19** | **39** | 10.313 | 69.120 | 74.179 | -5.059 | **73** | **181** | 12.663 | 216.622 | 232.105 | -15.483 |
| **20** | **40** | 10.338 | 72.348 | 77.909 | -5.561 | **74** | **184** | 12.538 | 222.562 | 237.628 | -15.067 |
| **22** | **48** | 10.538 | 78.124 | 84.083 | -5.959 | **74** | **186** | 12.513 | 223.696 | 238.109 | -14.413 |
| **23** | **51** | 10.563 | 81.607 | 87.689 | -6.081 | **79** | **197** | 12.438 | 239.414 | 255.739 | -16.325 |
| **24** | **52** | 10.588 | 85.021 | 91.286 | -6.265 | **82** | **206** | 12.663 | 244.513 | 260.696 | -16.184 |
| **25** | **55** | 10.638 | 88.005 | 94.641 | -6.636 | **82** | **207** | 12.663 | 244.245 | 260.698 | -16.453 |
| **26** | **54** | 10.563 | 92.093 | 99.120 | -7.027 | **82** | **208** | 12.688 | 243.968 | 260.182 | -16.215 |
| **26** | **56** | 10.738 | 90.570 | 97.507 | -6.936 | **83** | **209** | 12.638 | 248.781 | 264.404 | -15.623 |
| **26** | **58** | 10.763 | 90.493 | 97.284 | -6.791 | **90** | **232** | 12.863 | 263.790 | 281.711 | -17.921 |
| **27** | **59** | 10.763 | 94.008 | 101.026 | -7.019 | **92** | **238** | 12.988 | 266.822 | 285.169 | -18.347 |
| **28** | **58** | 10.638 | 98.749 | 105.999 | -7.250 | **93** | **239** | 13.213 | 265.958 | 283.348 | -17.390 |
| **28** | **60** | 10.713 | 98.258 | 105.229 | -6.970 | **94** | **239** | 13.213 | 269.005 | 286.395 | -17.390 |
| **28** | **61** | 10.738 | 97.932 | 105.002 | -7.069 | **95** | **243** | 13.238 | 271.068 | 288.886 | -17.818 |
| **28** | **62** | 10.763 | 97.693 | 104.753 | -7.060 | **96** | **245** | 13.263 | 273.681 | 291.441 | -17.760 |
| **28** | **64** | 10.813 | 97.352 | 104.265 | -6.913 | **97** | **247** | 13.288 | 276.222 | 293.917 | -17.694 |
| **29** | **63** | 10.863 | 99.768 | 107.502 | -7.734 | **98** | **249** | 13.288 | 278.760 | 296.940 | -18.180 |
| **29** | **65** | 10.913 | 99.392 | 107.010 | -7.618 | **99** | **254** | 13.338 | 280.580 | 298.856 | -18.277 |
| **30** | **64** | 10.888 | 102.982 | 110.950 | -7.968 | **100** | **253** | 13.313 | 283.825 | 302.422 | -18.598 |
| **30** | **66** | 10.963 | 102.435 | 110.197 | -7.762 | **101** | **255** | 13.338 | 286.351 | 304.870 | -18.519 |
| **30** | **68** | 10.963 | 102.566 | 110.191 | -7.625 | **102** | **255** | 13.313 | 289.370 | 308.466 | -19.097 |
| **30** | **70** | 11.063 | 101.578 | 109.192 | -7.615 | **103** | **257** | 13.338 | 291.896 | 310.910 | -19.014 |
| **38** | **88** | 11.163 | 128.258 | 137.088 | -8.830 | **104** | **261** | 13.388 | 293.915 | 312.757 | -18.842 |
| **39** | **89** | 11.263 | 130.137 | 139.445 | -9.309 | **105** | **262** | 13.388 | 296.673 | 315.761 | -19.088 |
| **41** | **93** | 11.313 | 135.766 | 145.942 | -10.177 | **106** | **263** | 13.388 | 299.454 | 318.791 | -19.338 |
| **48** | **110** | 11.588 | 155.872 | 166.795 | -10.923 | **107** | **262** | 13.363 | 302.695 | 322.376 | -19.681 |
| **48** | **112** | 11.638 | 155.307 | 166.081 | -10.774 | **108** | **265** | 13.388 | 304.958 | 324.799 | -19.841 |
| **48** | **114** | 11.663 | 154.748 | 165.715 | -10.967 | **109** | **266** | 13.388 | 307.684 | 327.777 | -20.094 |
| **48** | **116** | 11.688 | 154.543 | 165.361 | -10.818 | **110** | **269** | 13.413 | 309.938 | 330.186 | -20.248 |
| **49** | **115** | 11.513 | 160.253 | 171.382 | -11.129 | **111** | **272** | 13.438 | 312.155 | 332.551 | -20.396 |
| **50** | **112** | 11.638 | 161.289 | 172.989 | -11.700 | **112** | **277** | 13.488 | 313.866 | 334.291 | -20.425 |
| **50** | **116** | 11.688 | 160.771 | 172.245 | -11.474 | **114** | **289** | 13.613 | 316.836 | 337.146 | -20.310 |

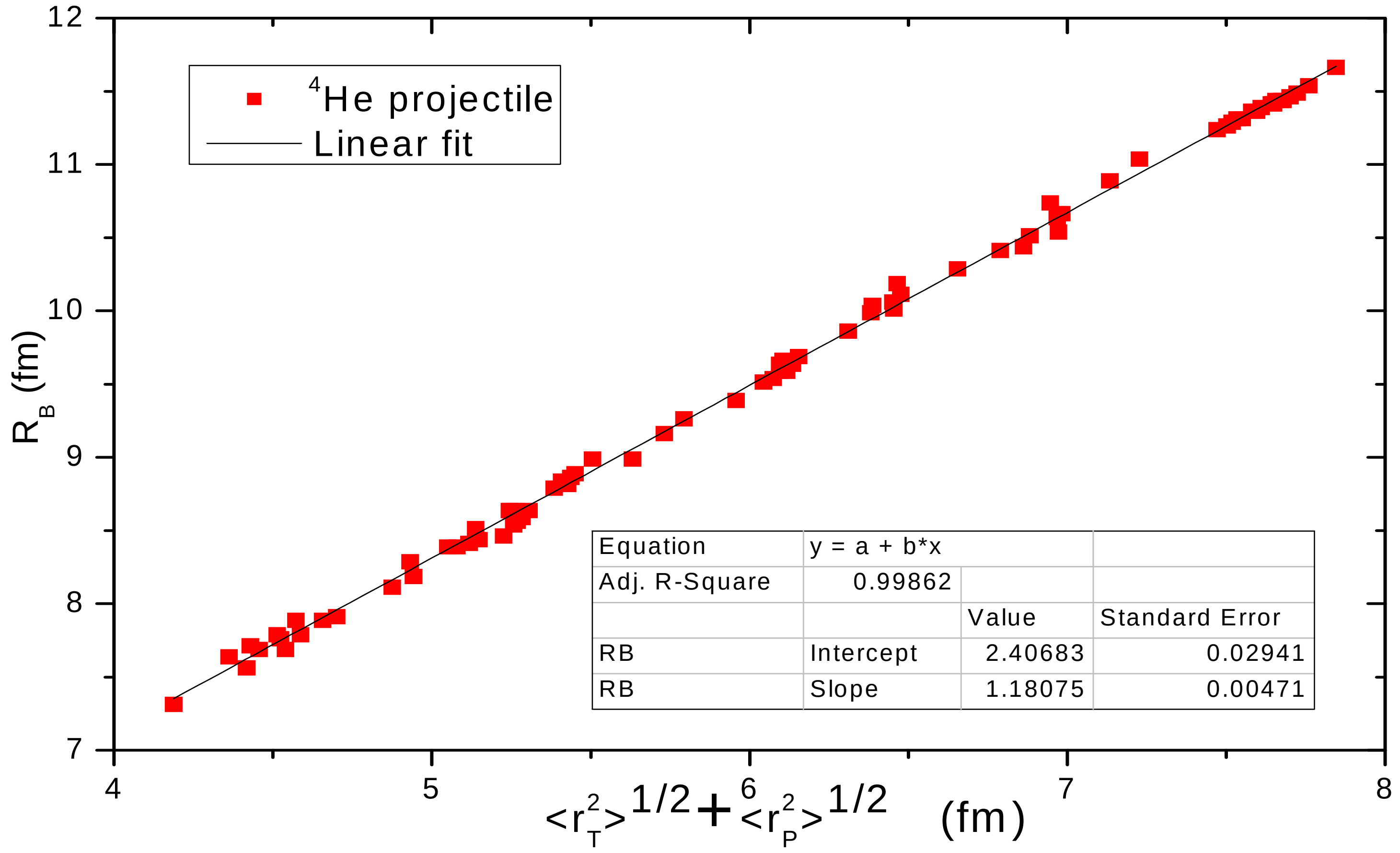

Figure (1.2-*a*) The variation of calculated barrier position ($R_B$) with $\langle r_T^2\rangle^{1/2} + \langle r_P^2\rangle^{1/2}$ for the reactions with $^4He$. The solid line is a linear fit to data and represented by

$$R_B = (1.18075 \pm 0.00471)\left(\langle r_T^2\rangle^{\frac{1}{2}} + \langle r_P^2\rangle^{\frac{1}{2}}\right) + (2.40683 \pm 0.02941)$$

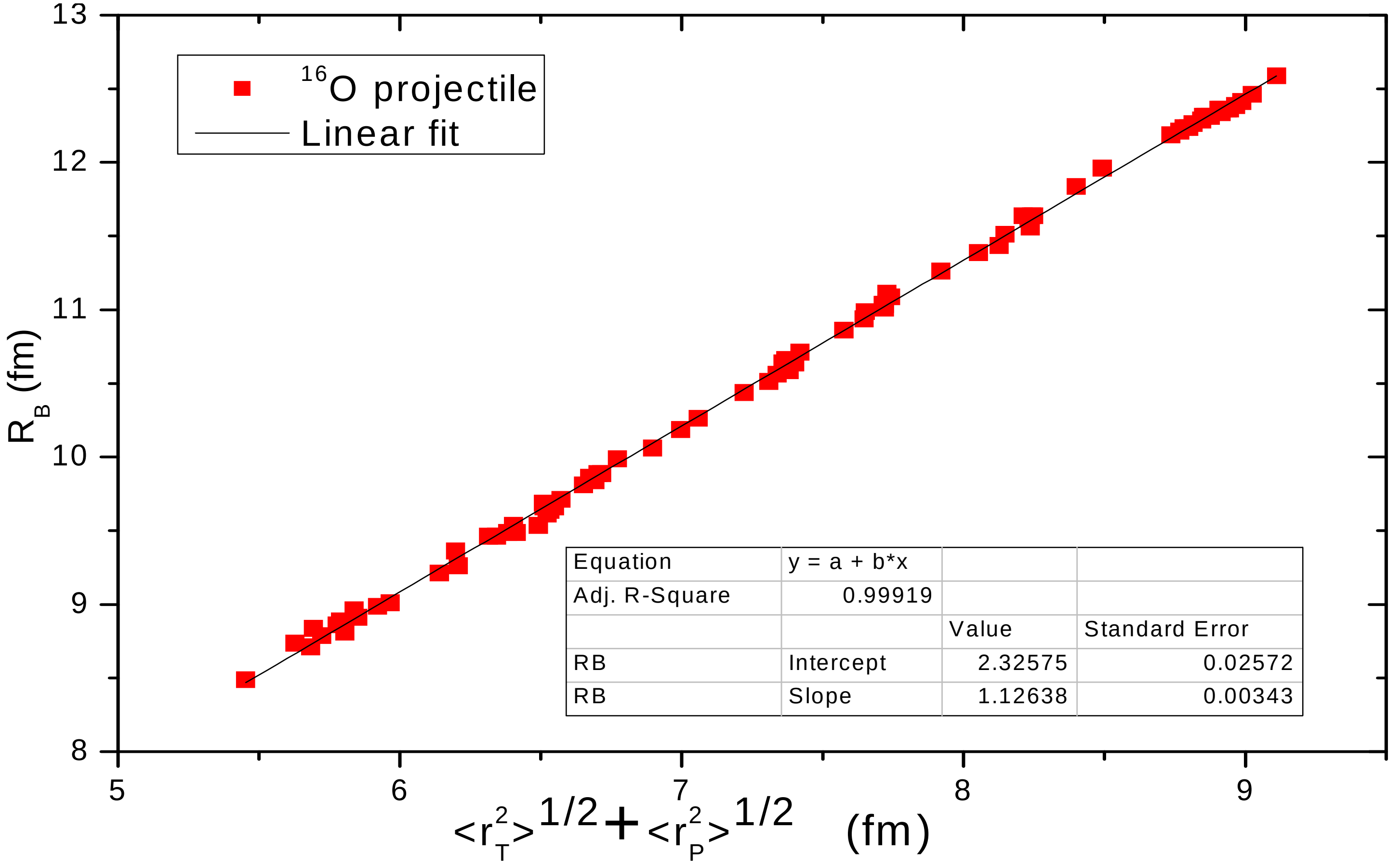

Figure (1.2-*b*) The same as Figure (1.2-*a*) but for $^{16}O$ as projectile. The solid line is a linear fit to data and represented by

$$R_B = (1.12638 \pm 0.00343)\left(\langle r_T^2\rangle^{\frac{1}{2}} + \langle r_P^2\rangle^{\frac{1}{2}}\right) + (2.32575 \pm 0.02572)$$

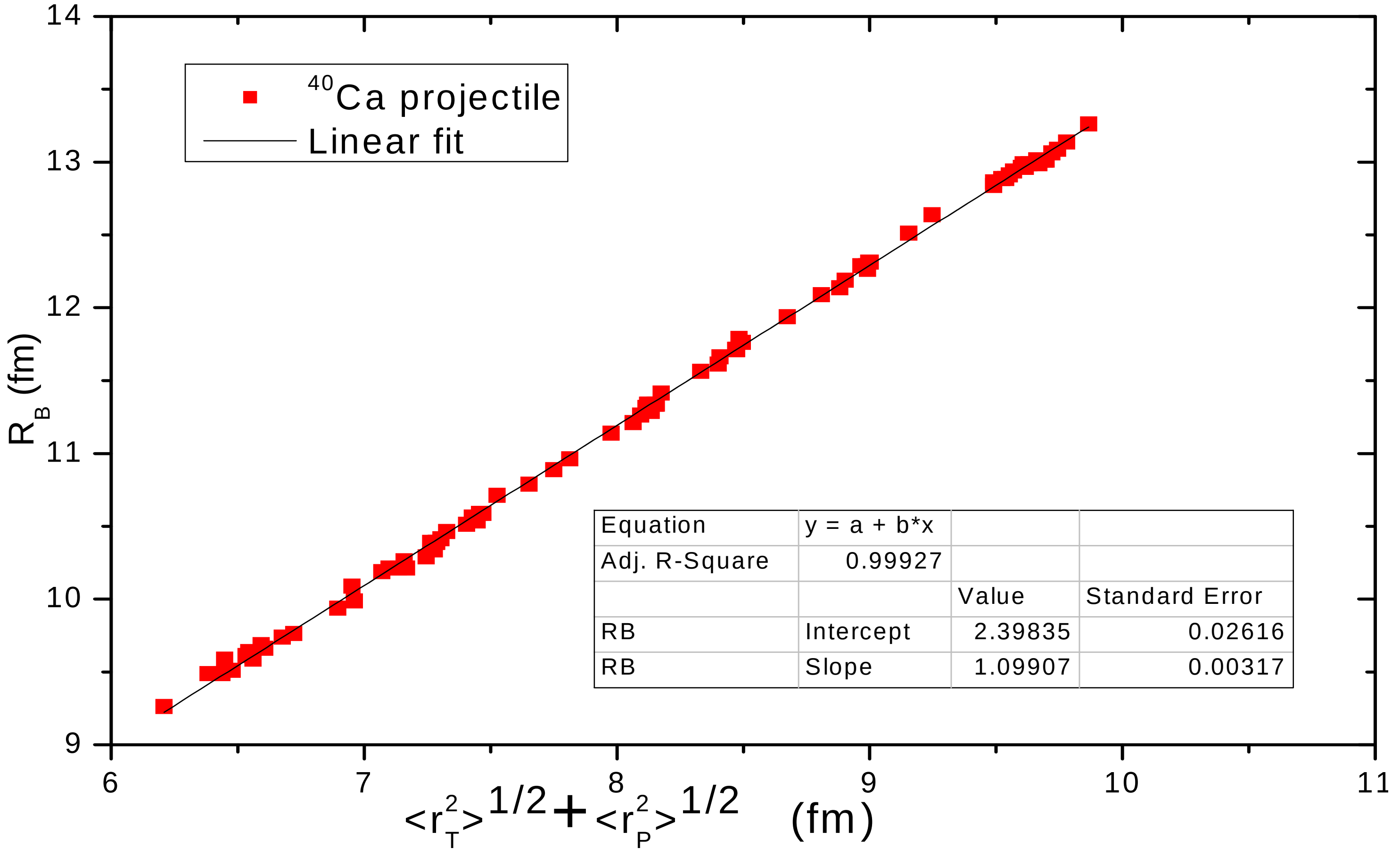

Figure (1.2-*c*) The same as Figure (1.2-*a*) but for $^{40}Ca$ as projectile. The solid line is a linear fit to data and represented by

$$R_B = (1.09907 \pm 0.00317)\left(\langle r_T^2\rangle^{\frac{1}{2}} + \langle r_P^2\rangle^{\frac{1}{2}}\right) + (2.39835 \pm 0.02616)$$

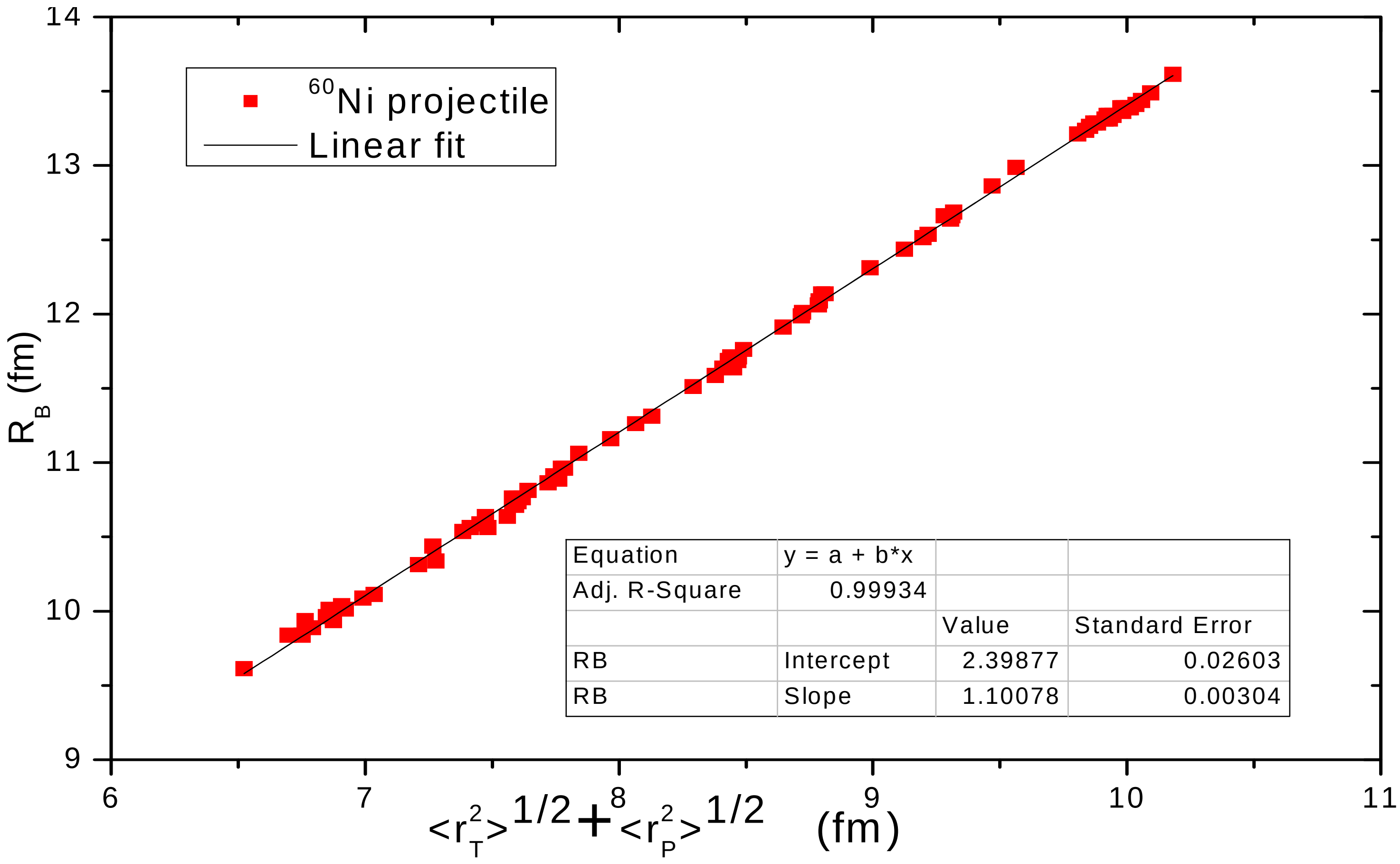

Figure (1.2-*d*) The same as Figure (1.2-*a*) but for $^{60}Ni$ as projectile. The solid line is a linear fit to data and represented by

$$R_B = (1.10078 \pm 0.00304)\left(\langle r_T^2\rangle^{\frac{1}{2}} + \langle r_P^2\rangle^{\frac{1}{2}}\right) + (2.39877 \pm 0.02603)$$

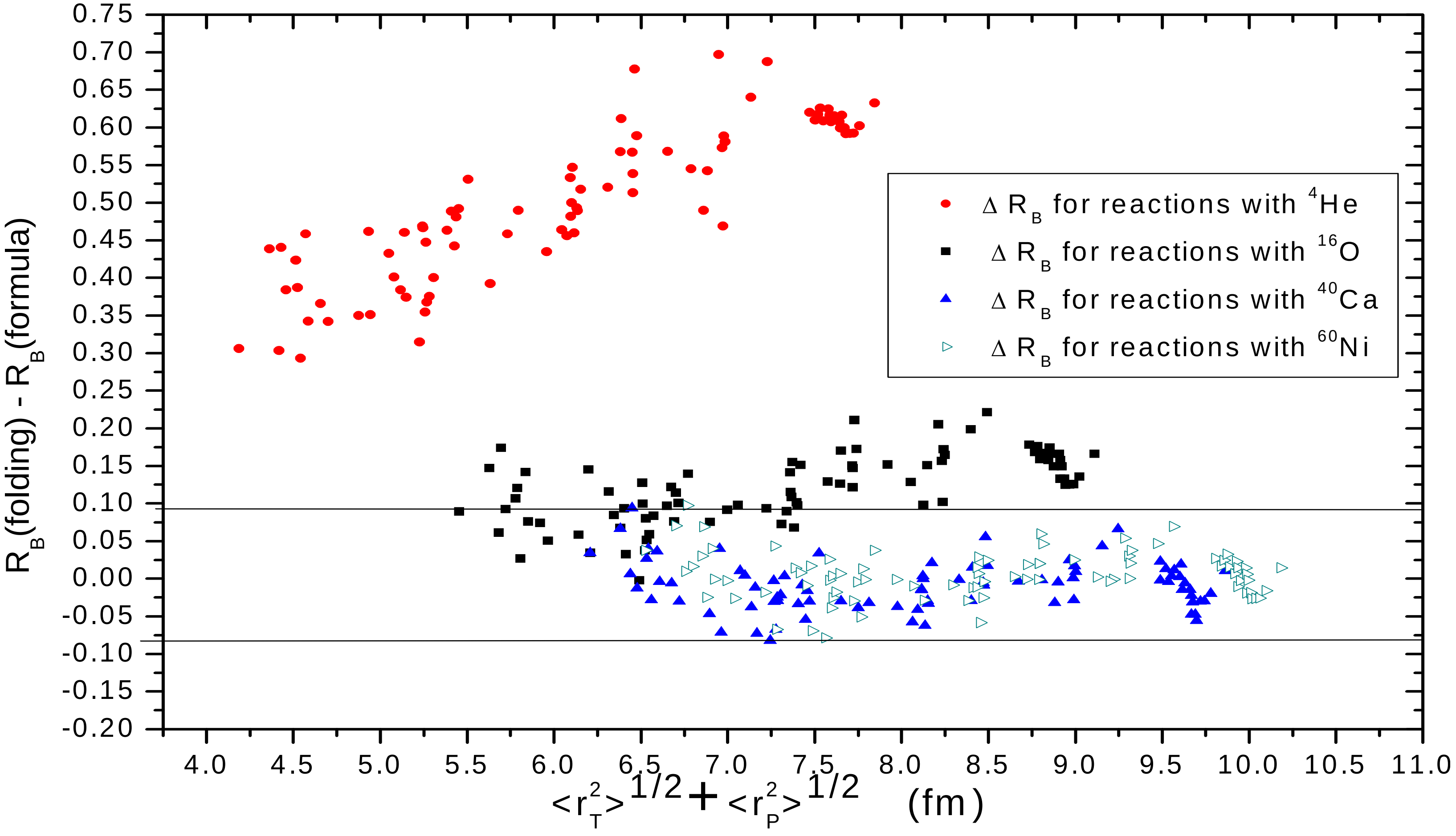

Figure (1.3-*a*) The differences between the values of $R_B$ calculated using DFM analysis and the values calculated using the formula $R_B = 1.1\left(\langle r_T^2\rangle^{\frac{1}{2}} + \langle r_P^2\rangle^{\frac{1}{2}}\right) + 2.4$.

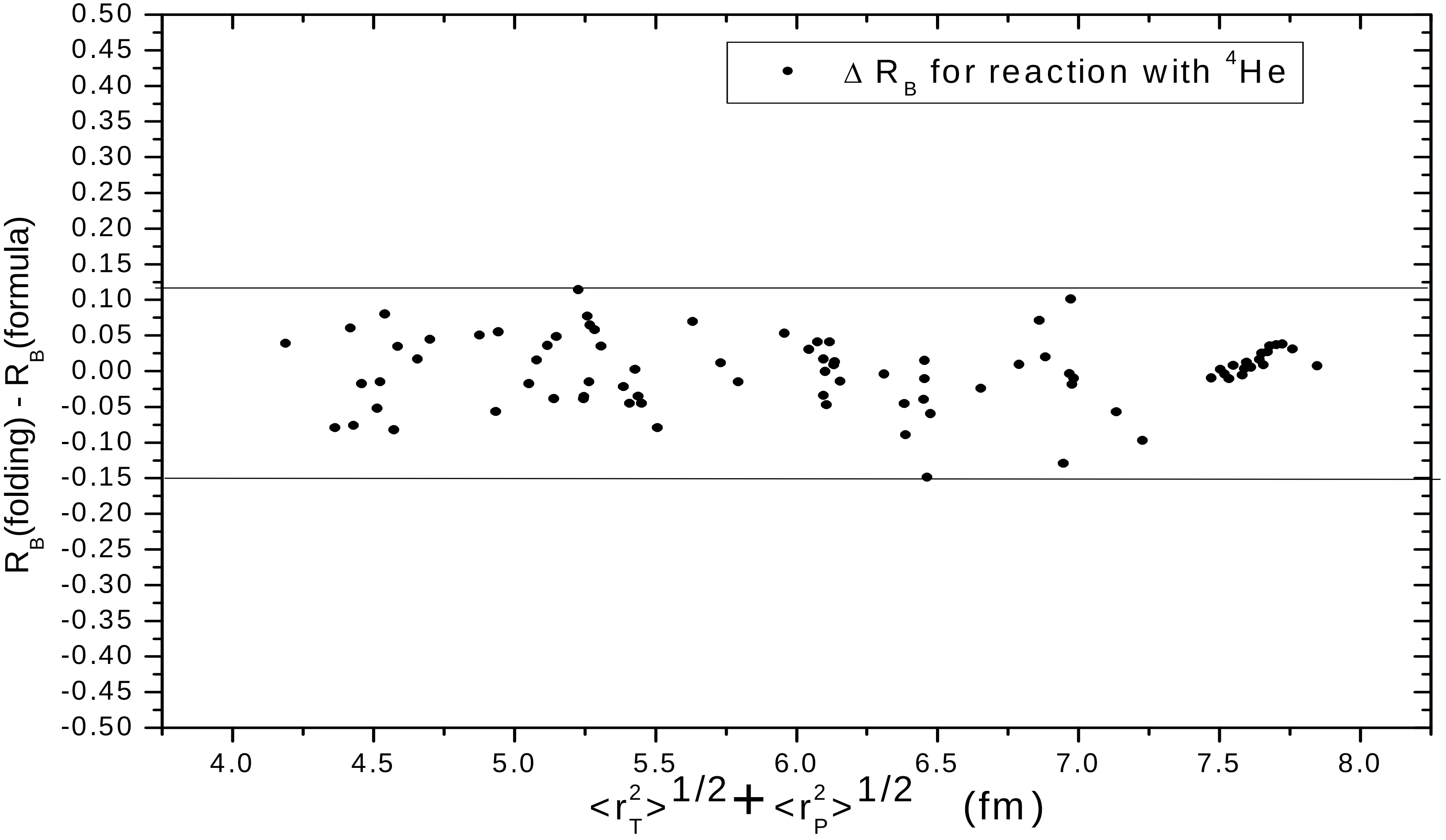

Figure (1.3-*b*) The differences between the values of $R_B$ calculated using DFM analysis and the values calculated using the formula $\left[R_B = 1.18075\left(\langle r_T^2\rangle^{\frac{1}{2}} + \langle r_P^2\rangle^{\frac{1}{2}}\right) + 2.40683\right]$, for reactions with $^{4}He$.

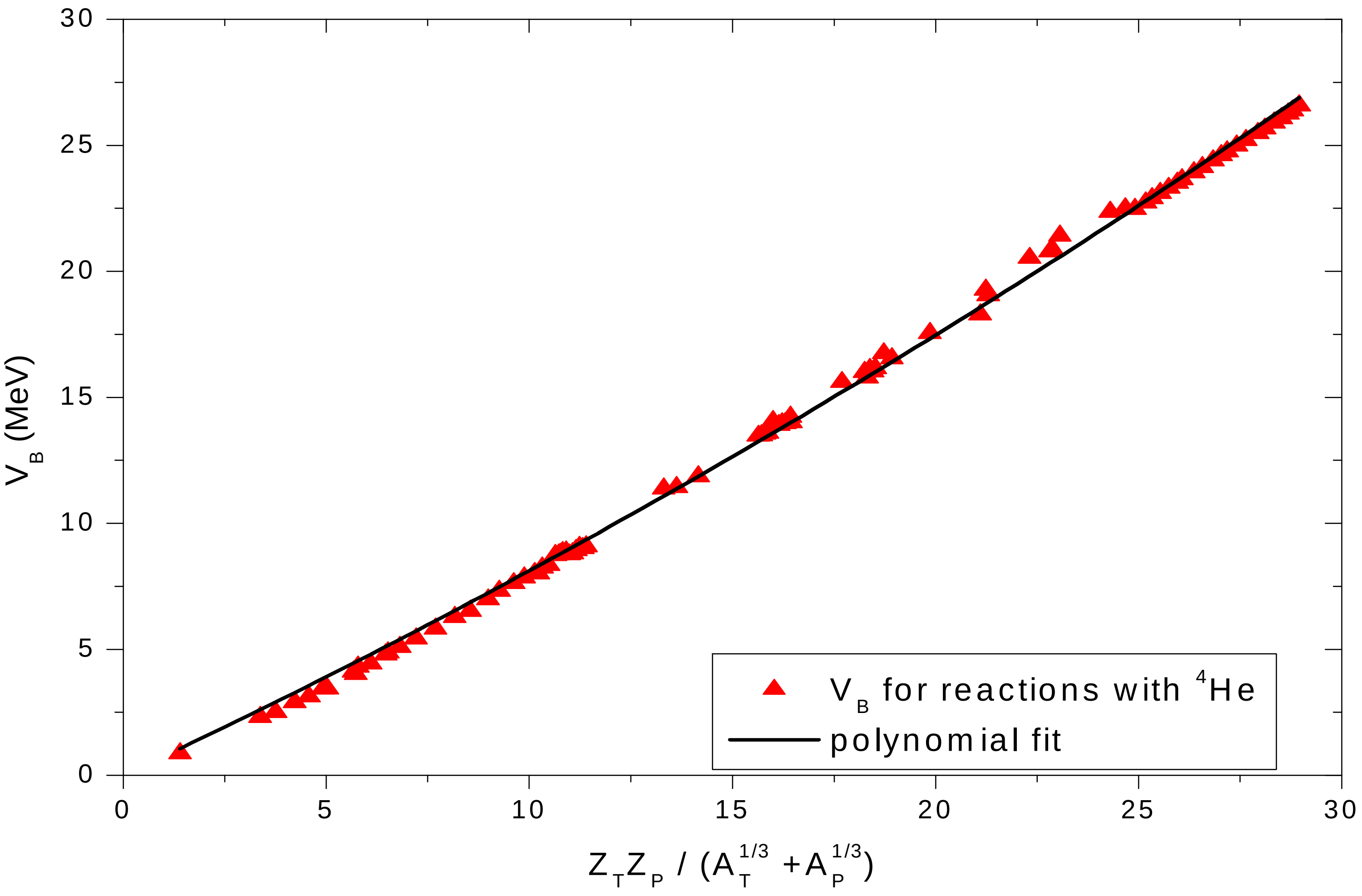

Figure (1.4-*a*) The variation of calculated barrier height ($V_B$) with $Z_1 Z_2/(A_T^{1/3} + A_P^{1/3})$ for the reactions with $^4He$. The solid line is a second order polynomial fit to data and represented by

$$V_B = (0.74999 \pm 0.00681)\,\frac{Z_T Z_P}{A_T^{1/3} + A_P^{1/3}} + (6.15 \pm 0.292177) \times 10^{-3} \left[\frac{Z_T Z_P}{A_T^{1/3} + A_P^{1/3}}\right]^2$$

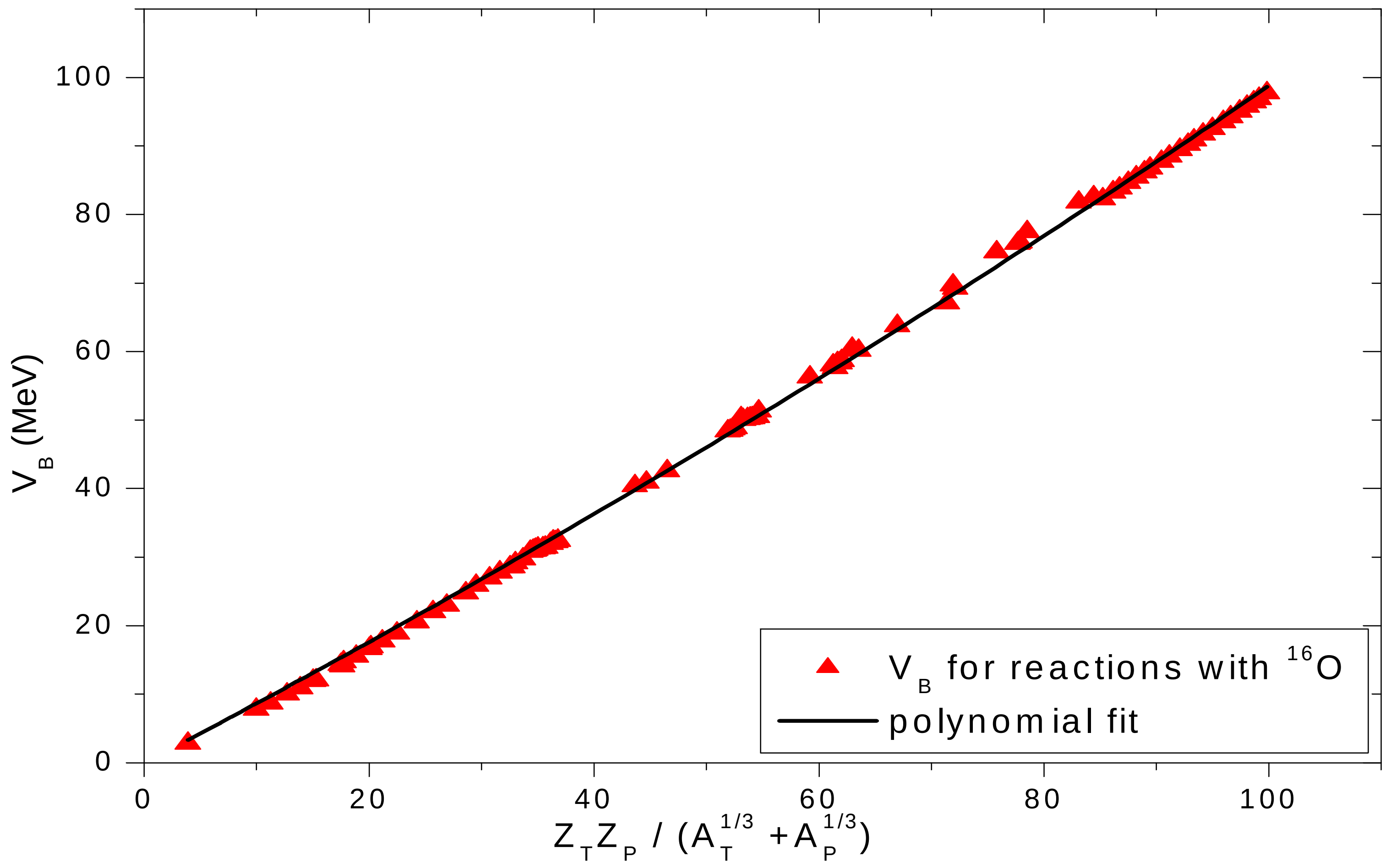

Figure (1.4-*b*) The same as Figure (1.4-*a*) but for $^{16}O$ as projectile. The solid line is a second order polynomial fit to data and represented by

$$V_B = (0.8529 \pm 0.00561)\,\frac{Z_T Z_P}{A_T^{1/3} + A_P^{1/3}} + (13.5 \pm 0.69832) \times 10^{-4} \left[\frac{Z_T Z_P}{A_T^{1/3} + A_P^{1/3}}\right]^2$$

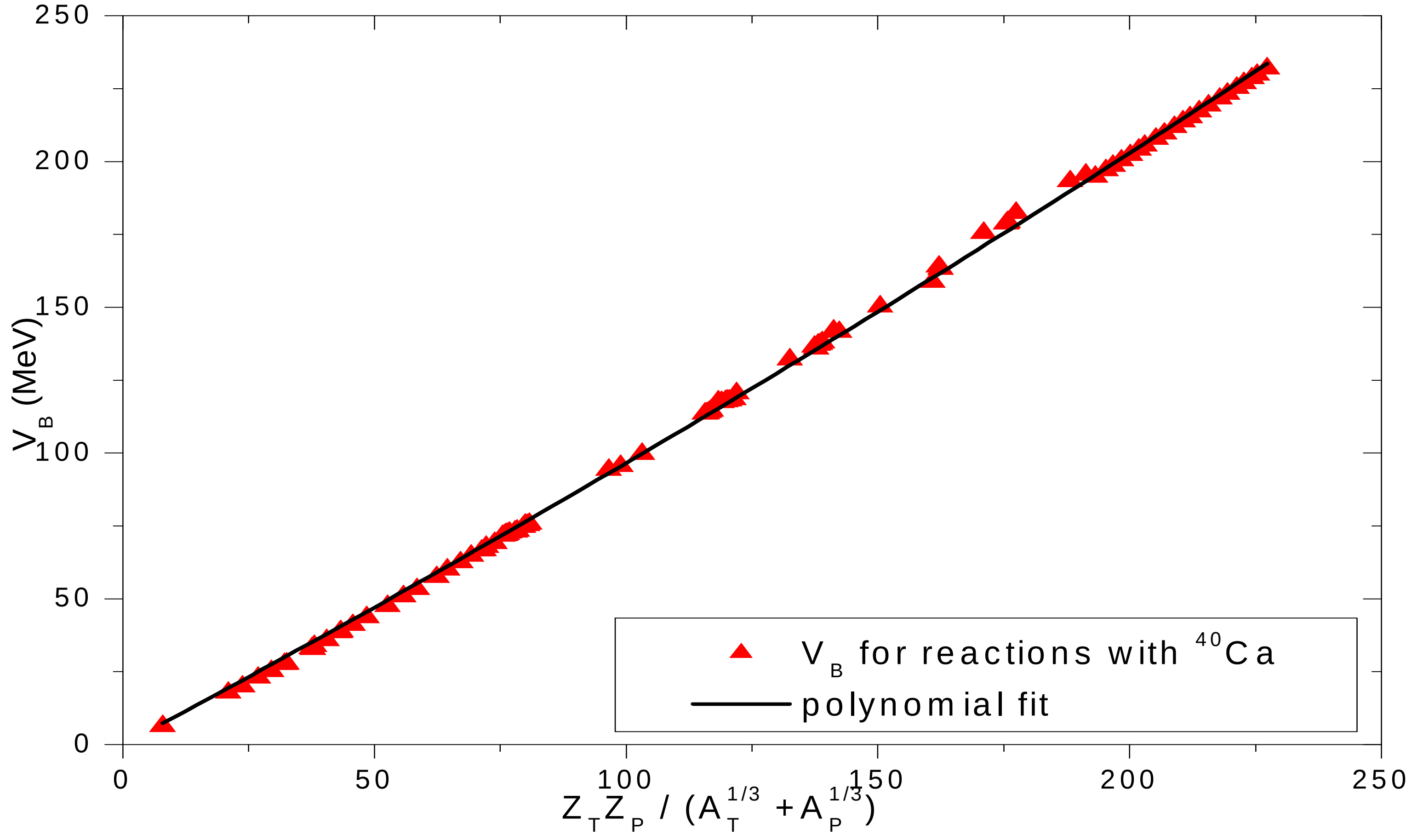

Figure (1.4-*c*) The same as Figure (1.4-*a*) but for $^{40}Ca$ as projectile. The solid line is a second order polynomial fit to data and represented by

$$V_B = (0.91656 \pm 0.00503)\,\frac{Z_T Z_P}{A_T^{1/3} + A_P^{1/3}} + (4.89575 \pm 0.275422) \times 10^{-4}\left[\frac{Z_T Z_P}{A_T^{1/3} + A_P^{1/3}}\right]^2$$

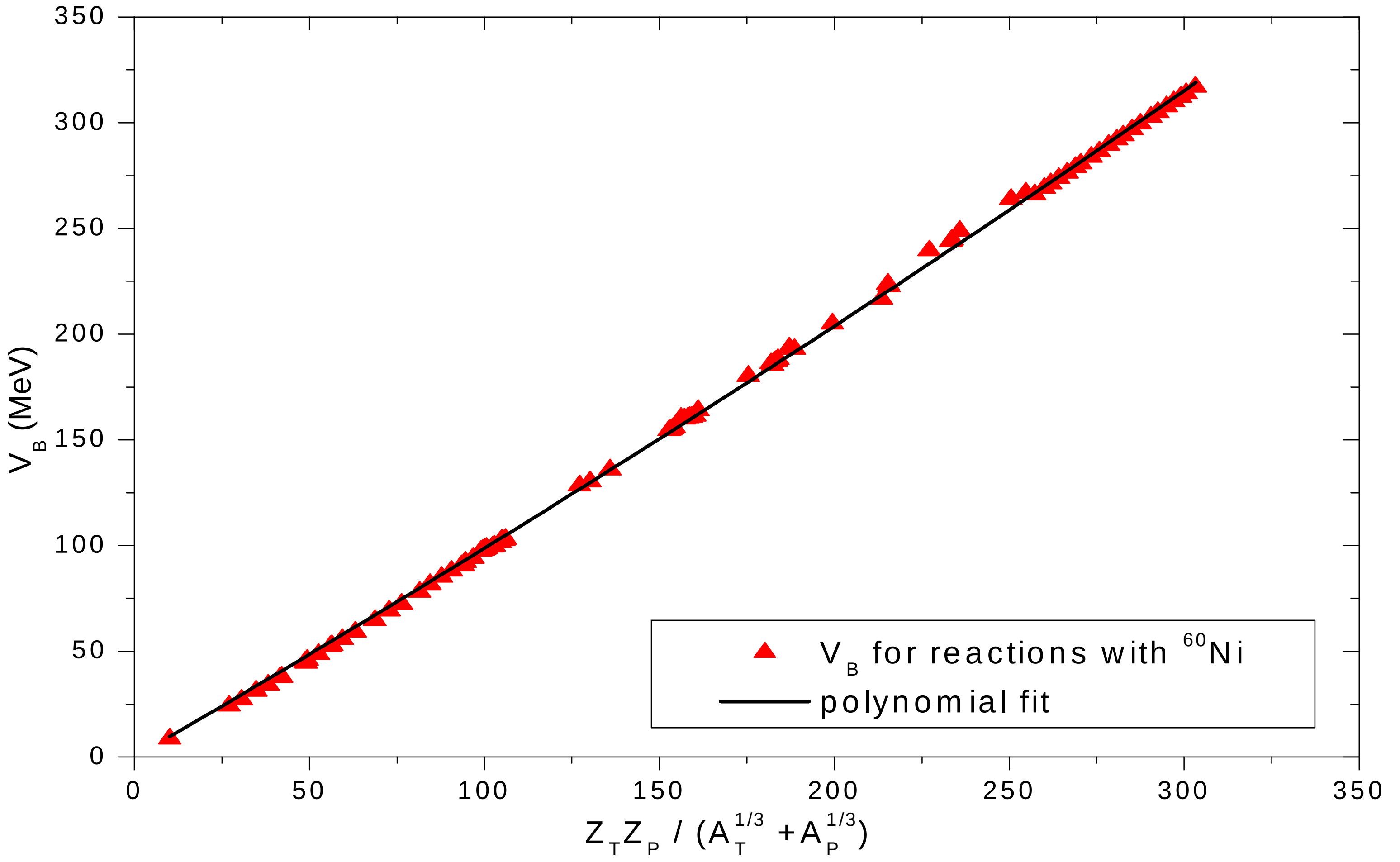

Figure (1.4-*d*) The same as Figure (1.4-*a*) but for $^{60}Ni$ as projectile. The solid line is a second order polynomial fit to data and represented by

$$V_B = (0.95546 \pm 0.00477)\,\frac{Z_T Z_P}{A_T^{1/3} + A_P^{1/3}} + (3.17332 \pm 0.195828) \times 10^{-4} \left[\frac{Z_T Z_P}{A_T^{1/3} + A_P^{1/3}}\right]^2$$

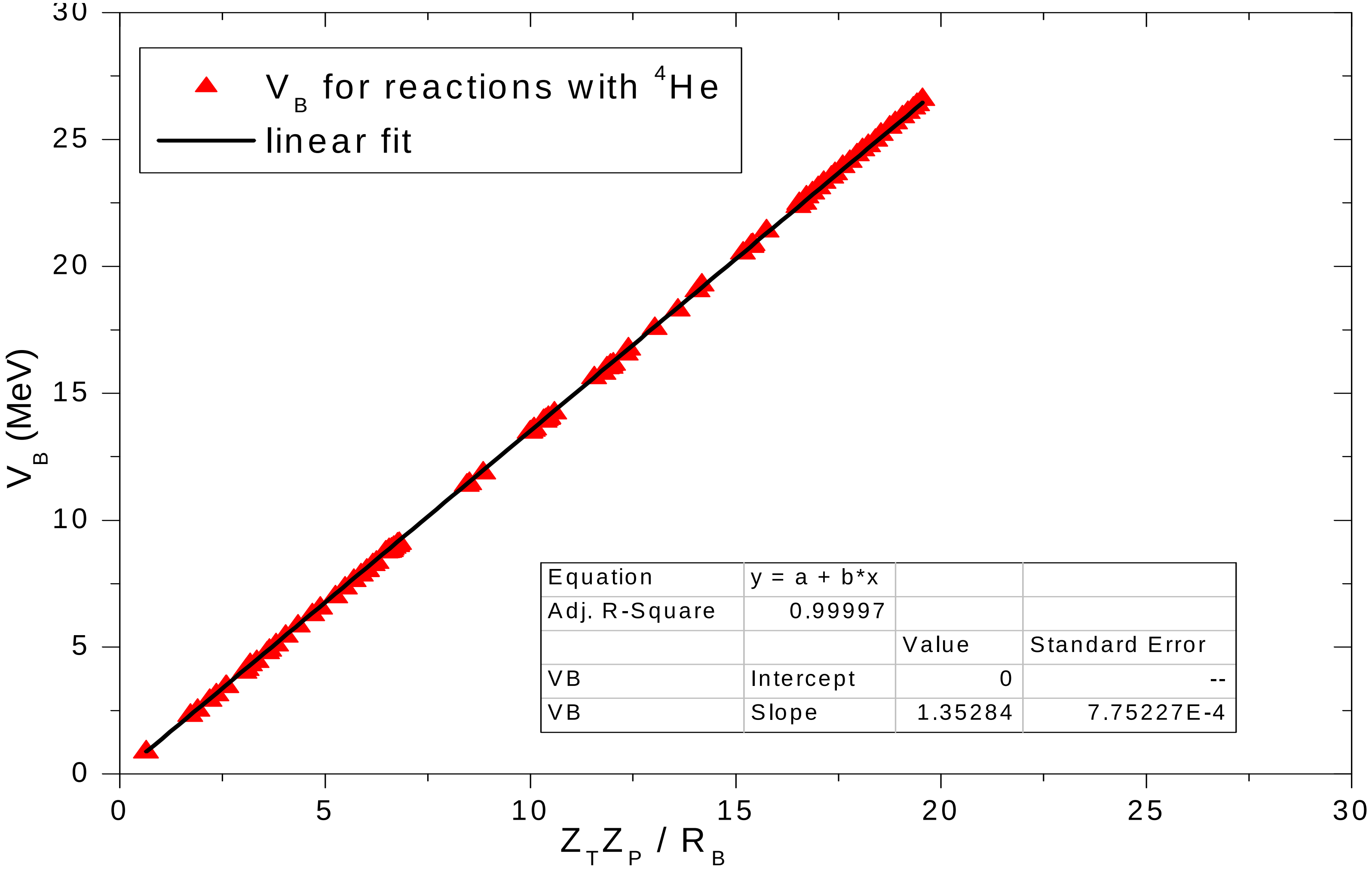

Figure (1.5-*a*) The variation of calculated barrier height ($V_B$) with ($Z_T Z_P/R_B$) for the reactions with $^4He$. The solid line is a linear fit to data and represented by

$$V_B = (1.35284 \pm 7.75227 \times 10^{-4}) \times Z_T Z_P / R_B \, .$$

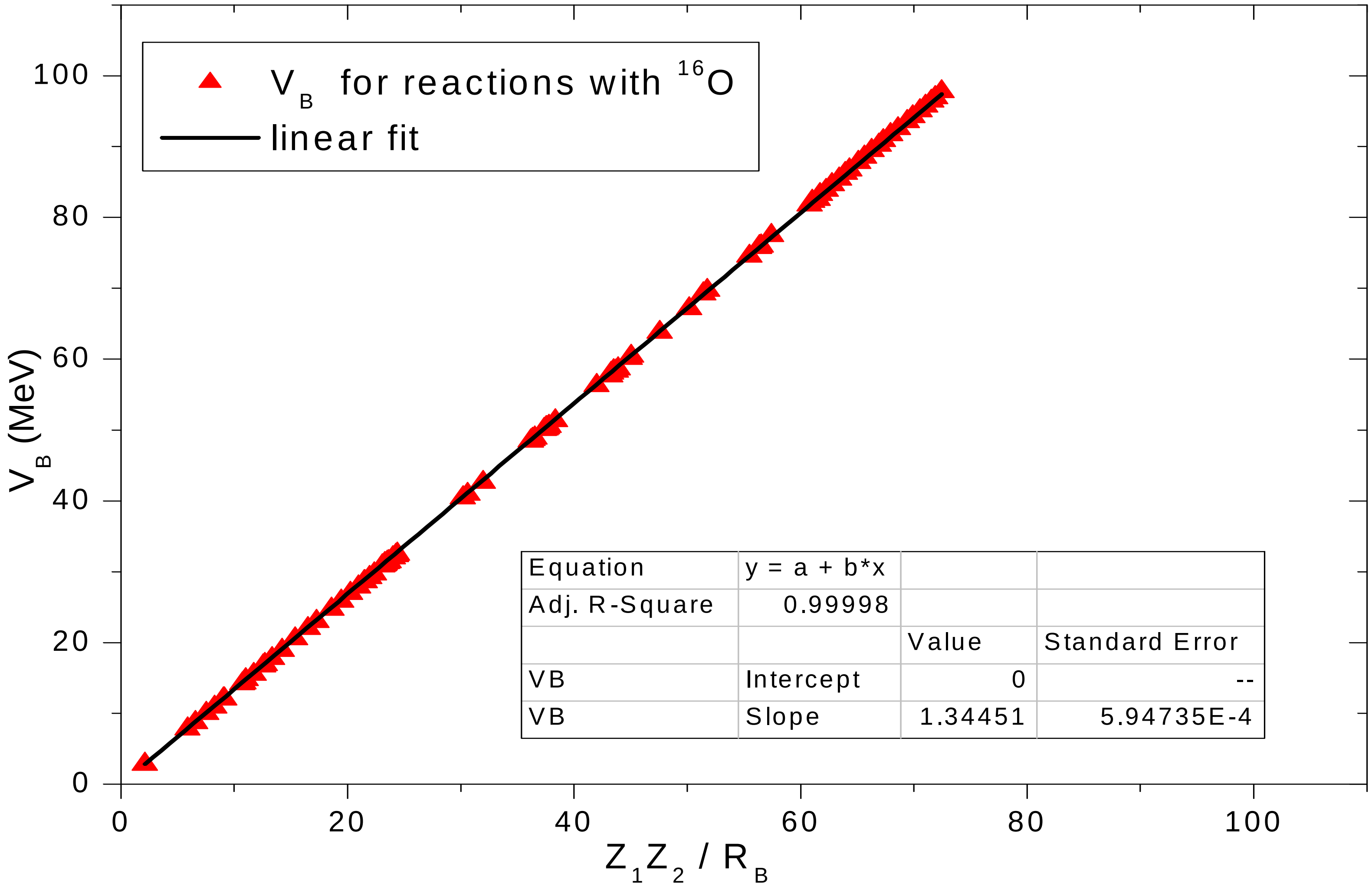

Figure (*1.5-b*) The same as Figure (1.5-*a*) but for ${}^{16}O$ projectile. The solid line is a linear fit to data and represented by

$$V_B = (1.34451 \pm 5.94735 \times 10^{-4}) \times Z_T Z_P / R_B\,.$$

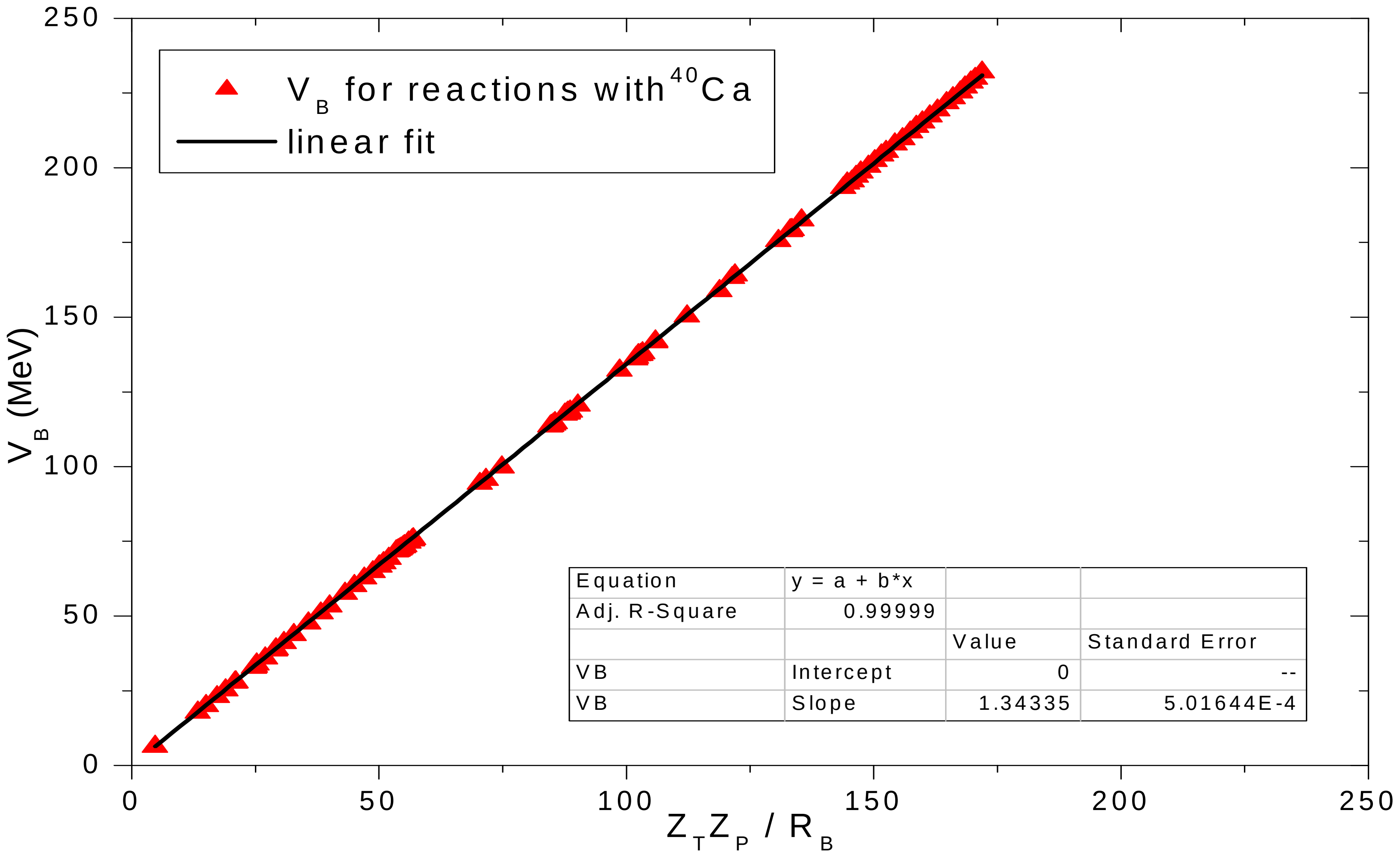

Figure (1.5-*c*) The same as Figure (1.5-*a*) but for $^{40}Ca$ projectile. The solid line is a linear fit to data and represented by

$$V_B = (1.34335 \pm 5.01644 \times 10^{-4}) \times Z_T Z_P / R_B \,.$$

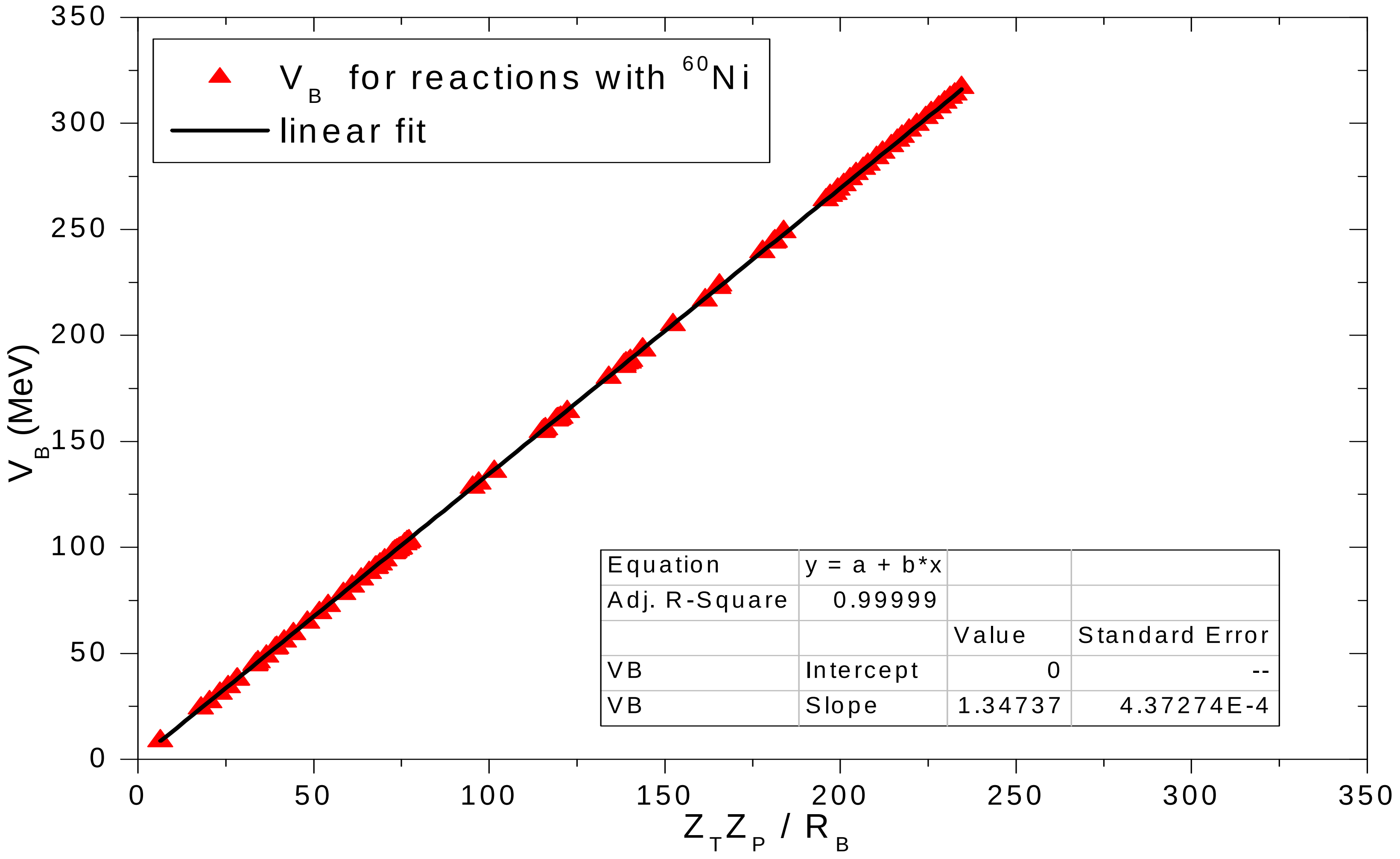

Figure (1.5-*d*) The same as Figure (1.5-*a*) but for $^{60}Ni$.projectile. The solid line is a linear fit to data and represented by

$$V_B = (1.34737 \pm 4.37274 \times 10^{-4}) \times Z_T Z_P / R_B \, .$$

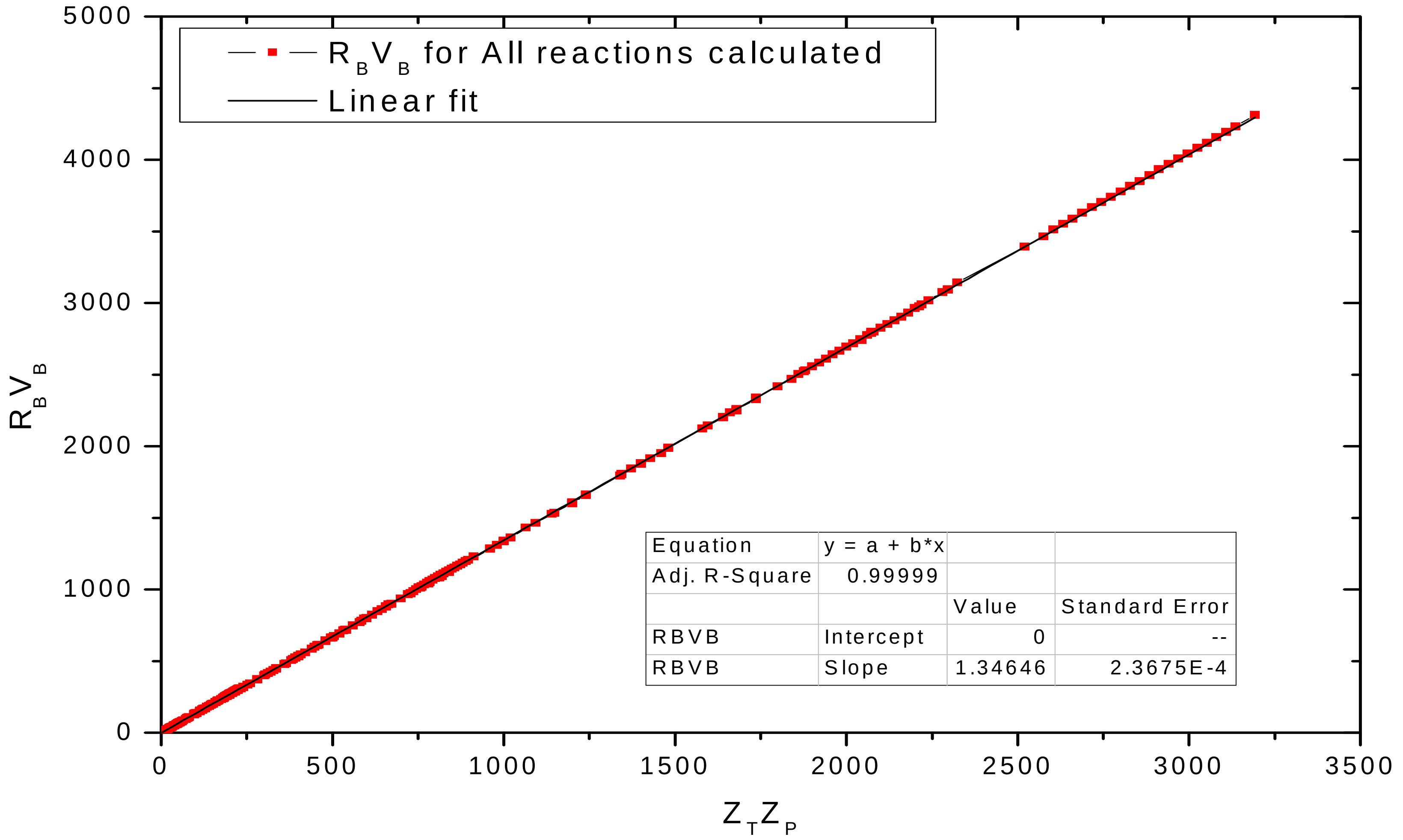

Figure (1.6-*a*) The behavior of ($R_B\,V_B$) product for all reaction done within the present chapter with $Z_TZ_P$. The solid line is a linear fit to data and represented by

$$R_B \times V_B = (1.34646 \pm 2.3675 \times 10^{-4}) \times Z_T Z_P.$$

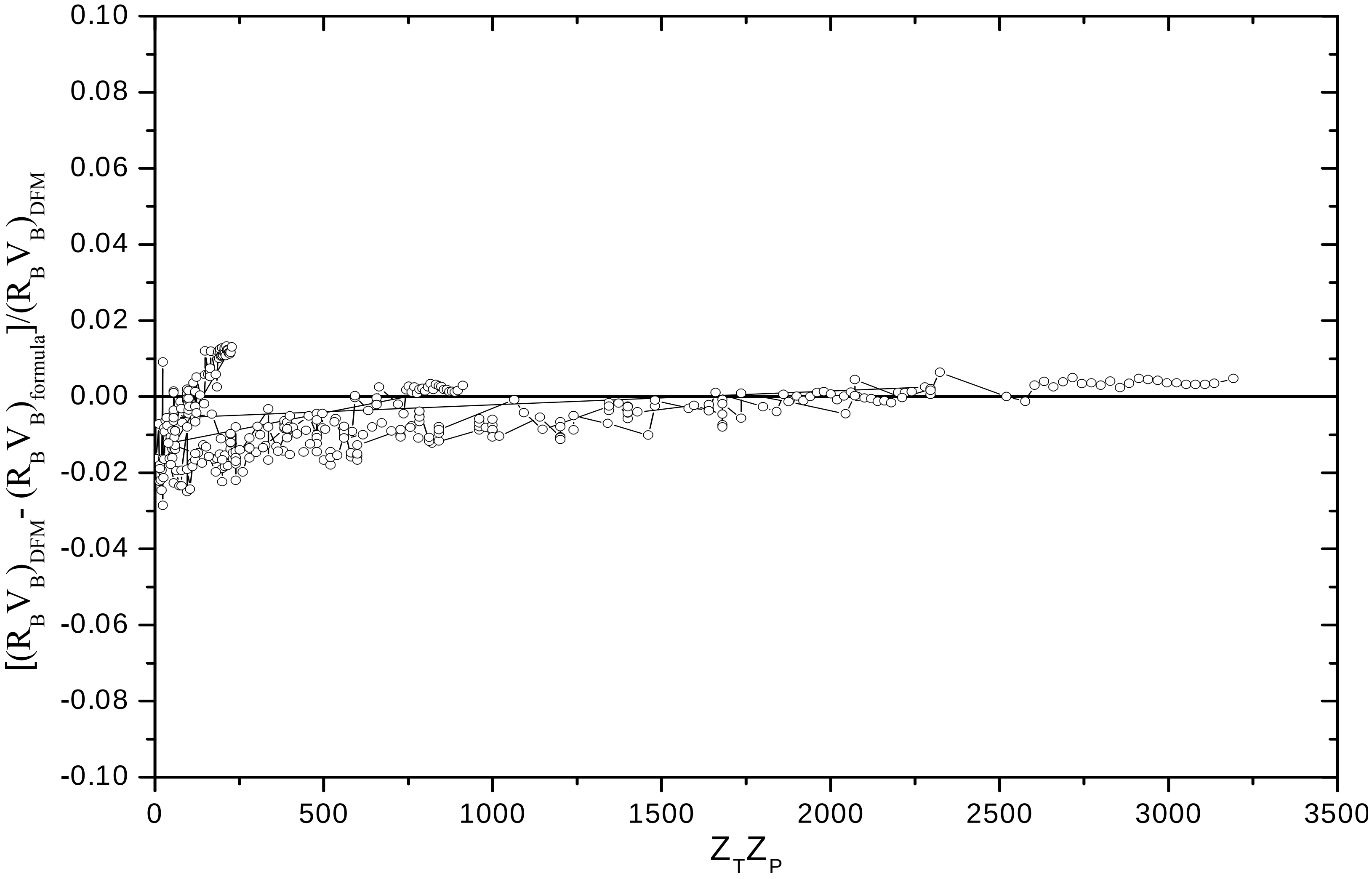

Figure (1.6-b) the relative differences in the values of $R_B\,V_B$ calculated using DFM and the values calculated using formula (1.13) for all reactions studied in chapter (1).

## 1.5 Discussion

In the present study of the systematic behaviour of the fusion barrier parameters, we considered $^{4}He$, $^{16}O$, $^{40}Ca$, and $^{60}Ni$ as projectiles and a large number of different target nuclei. We extended the study to a number of SHE with $Z \leq 114$. The DFM calculations were performed to find the fusion barrier parameters, $V_B$ and $R_B$, as well as the values of nuclear and Coulomb potentials at $R = R_B$, for reactions of $^{4}He$, $^{16}O$, $^{40}Ca$, and $^{60}Ni$ with large number of nuclei as given in tables (1.2-*a*, *b*, *c*, and *d*). From the values of $V_C$ and $V_N$ in tables (1.2) it is clear that major contribution to $V_B$ comes from Coulomb potential, and the magnitude of nuclear potential at $R_B$ is less than 8% of the magnitude of Coulomb potential at $R_B$.

Some values of $R_B$ show unexpected behaviour, for example the reaction between $^{16}O$ and $^{25}Mg$ has value of $R_B$ =8.963 *fm*, while the reaction between $^{16}O$ and $^{26}Mg$ has $R_B$ =8.888 *fm*, in this case the mass number (*A*) increases, while $R_B$ decreases. Similar behaviour is found for reactions of $^{16}O$ with ($^{148}Nd$ and $^{150}Nd$) and ($^{181}Ta$, $^{184}W$). This behaviour shows that $R_B$ is not sensitive to the mass only, but also to the density distribution. From table (1.1) and comparing density distribution parameters for the examples mentioned above we note that the nuclei which have large diffuseness parameters, give unexpected large fusion radii. For $^{25}Mg$ and $^{26}Mg$, the diffuseness values are (0.608 *fm*) and (0.524 *fm*) respectively, and $^{181}Ta$ has diffuseness parameter of 0.64 *fm*. Moreover, reactions with target nuclei such as $^{186}W$ and $^{209}Bi$, which have relatively small diffuseness parameters, (0.480 *fm* and 0.468 *fm* respectively), provide smaller values of $R_B$ than reactions with the lighter nuclei $^{184}W$ and $^{208}Pb$ respectively. Large value of diffuseness enhances the attractive nuclear contribution before and at the position of Coulomb barrier, and hence shifts the barrier outwards.

In a first trial we represented $R_B$ as a function of $\mathrm{A_1}^{1/3} + \mathrm{A_2}^{1/3}$ -as is usually done [51, 54, 55] - but the dependence, in this case, was not regular. More regular dependence of $R_B$ is obtained with the sum of root mean square (rms) radii of interacting nuclei. Figures (1.2-a, b, c, and d) show the variation of $R_B$ parameter with the sum of the calculated rms radii of the interacting nuclei. The rms radius is calculated from the equation

$$< r^2 >^{\frac{1}{2}} = \left[\frac{\int_0^\infty \rho(r) r^4 \, dr}{\int_0^\infty \rho(r) r^2 \, dr}\right]^{\frac{1}{2}}$$

where $\rho$(r) is the density distribution of the nucleus described in section (1.3). As shown in figures (1.2-*a*, *b*, *c*, and *d*) the variation of $R_B$ with sum of rms radii of colliding nuclei shows systematic variation with very little spread of data, the straight–line fit of the data in those figures clearly shows the success of first order parameterization of $R_B$ in terms of rms radii. The linear fit in figures (1.2-*a*, *b*, *c*, and *d*) has a general form of

$$R_B = B\left(\langle r_T^2\rangle^{\frac{1}{2}} + \langle r_P^2\rangle^{\frac{1}{2}}\right) + C.$$

This is a simple first order expression with two coefficients *B* and *C*, the values of the coefficients are given in figures (1.2-*a*, *b*, *c*, and *d*) for reactions of deferent targets with $^{4}He$, $^{16}O$, $^{40}Ca$, and $^{60}Ni$ respectively. In case of $^{40}Ca$, and $^{60}Ni$ the coefficients are almost equal, $B \cong 1.1$ and $C \cong 2.4\ fm$. The value of *B* for the reactions involving $^{4}He$ is larger than the value mentioned above by about (7.3%) and for the reactions with $^{16}O$ is larger by about (2.4%). The second coefficient $R_0$ for reactions with $^{4}He$ has the same value as for $^{40}Ca$, and $^{60}Ni$ but for reactions with $^{16}O$ is smaller by about (3%). Small differences in the values of *B* and $R_0$ for reactions with light nuclei prove the possibility of using a single mathematical expression for the potential barrier position ($R_B$) between heavy ion (HI) pairs with high accuracy. So that, the expression of $R_B$ becomes,

$$R_B = 1.1\left(\langle r_T^2\rangle^{\frac{1}{2}} + \langle r_P^2\rangle^{\frac{1}{2}}\right) + 2.4, \tag{1.12}$$

this expression gives a simple and direct method to calculate the potential barrier radius ($R_B$) at least for large number of HI reactions from the knowledge of rms radii of reaction pair. The accuracy of the above formula is tested by comparing the values of $R_B$ calculated using DFM and the values calculated using formula (1.12), the differences in values are shown in figure (1.3-*a*) for the reactions with $^{4}He$, $^{16}O$, $^{40}Ca$, and $^{60}Ni$. For the reactions with $^{40}Ca$, and $^{60}Ni$ all deviations are located between the two horizontal lines -0.08 *fm* and 0.09 *fm* and then the values of $R_B$ for these reactions can be calculated using the above formula within about 0.1 *fm*. Deviations for reactions with $^{16}O$ reach to 0.22 *fm* and for reactions with $^{4}He$ reach to 0.72 *fm*. Deviations for reactions with $^{4}He$ is unacceptable because of thier large values, and because of location of all deviations above 0.3 *fm i.e.* the most accurate result has an error of 0.3 *fm*. Figure (1.3-*b*) is the same as Figure (1.3-*a*) but $R_B$ is calculated using a separate formula, other than equation (1.12), obtained for reactions of $^{4}He$ with deferent targets. This formula is

$$R_B = 1.18075\left(\langle r_T^2\rangle^{\frac{1}{2}} + \langle r_P^2\rangle^{\frac{1}{2}}\right) + 2.40683$$

In this case the deviations are well distributed around the zero deviation line, and all deviations are located between the two horizontal lines -0.15 *fm* and 0.12 *fm*. Thus, equation (1.12) can be used satisfactory for reactions involving projectiles of mass numbers $A_p \geq 16$, while for $^4He$ as projectile, $R_B$ can be obtained from the formula mentioned above.

As mentioned before, the major contribution to $V_B$ comes from Coulomb potential, so one can predict strong dependence of $V_B$ on relevant quantities such as $Z_T Z_P/[A_T^{1/3} + A_P^{1/3}]$ or $Z_T Z_P/[\langle r_T^2\rangle^{\frac{1}{2}} + \langle r_P^2\rangle^{\frac{1}{2}}]$. Authors in reference [56] have suggested a parameterization of $V_B$ for light colliding nuclei with mass numbers up to 64, as a second order function of $Z_T Z_P/[A_T^{1/3} + A_P^{1/3}]$ as

$$V_B = (0.845 \pm 0.02)\ \frac{Z_T Z_P}{A_T^{1/3} + A_P^{1/3}} + (1.3 \pm 0.25) \times 10^{-3}\left[\frac{Z_T Z_P}{A_T^{1/3} + A_P^{1/3}}\right]^2.$$

In fact this expression is simple and direct, but it is not general for all nuclei or needs to be generalized. In order to obtain similar general expression, figures (1.4-*a*, *b*, *c*, and *d*) show the variation of calculated barrier height ($V_B$) with the quantity $Z_T Z_P/[A_T^{1/3} + A_P^{1/3}]$, for reactions with projectiles $^{4}He$, $^{16}O$, $^{40}Ca$, and $^{60}Ni$ respectively and the results are fitted to a polynomial of degree two with no zero order term to get an expression similar to the one mentioned above. It is important to note that fits in figures (1.4-*a, b, c*, and *d*) are restricted to pass through the origin *i.e.* the zero order term is restricted to be zero. Without this restriction, fits will contain a negative zero order term which leads to unphysical result of negative barrier height. Figures (1.4-*a*, *b*, *c*, and *d*) show systematic variation with very little spread of points, and each graph can accurately represented by the expression obtained by fitting the data points. The four expressions obtained for reactions of different nuclei with $^{4}He$, $^{16}O$, $^{40}Ca$, and $^{60}Ni$ are similar but the coefficients are not equal. The differences between the coefficients for different projectiles can not be neglected. The coefficients of the first order terms are 0.74999, 0.8529, 0.91656, and 0.95546 for reactions with $^{4}He$, $^{16}O$, $^{40}Ca$, and $^{60}Ni$ respectively; it is clear that the values of this coefficient increase with irregular step and do not oscillate around a mean value as obtained in the parameterization of $R_B$ mentioned before. The coefficients of the second order terms are very small in magnitude compared to the coefficients of the first order terms but their values decrease in very large steps from figure (1.4-*a*) to figure (1.4-*d*). Comparisons between the four formulae show that it will not be easy to suggest one simple and general formula for $V_B$ as a function of $Z_T Z_P/[A_T^{1/3} + A_P^{1/3}]$.

The previous discussion shows the systematic dependence of the potential barrier parameters on some structural quantities of the interacting nuclei. Now, it is logical to discuss the relation - if it exists - between the potential barrier parameters each other. In normal or even extreme conditions, charge number ($Z$) and mass number ($A$) can not exceed definite regions of stability on the $A$-$Z$ plane, this fact supports the idea of systematic Interdependent variation of the potential barrier parameters, because the regular increase of $Z$ with $A$ indicates a regular growing of both repulsive and attractive fields of the nucleus. Figures (1.5-*a*, *b*, *c*, and *d*) show the variation of calculated barrier height ($V_B$) with ($Z_T Z_P/R_B$) for reactions with $^{4}He$, $^{16}O$, $^{40}Ca$, and $^{60}Ni$ respectively, and the results of linear fit to data, the graphs show very little scatter of data, moreover, the values of coefficients of the resulting linear fits are close to each other, and an average coefficient may be deduced to find a relation between the potential barrier parameters ($V_B$ and $R_B$) as *"the product of the potential barrier parameters ($V_B$ and $R_B$) is directly proportional to* $Z_T Z_P$*"*, this relation can be stated symbolically as

$$R_B \times V_B = C \times Z_T Z_P,$$

where the constant of proportionality ($C$) is the same as the coefficient of resulting linear fit in figures (1.5-*a*, *b*, *c*, and *d*), and has a value around 1.345. This behavior is illustrated in figure (1.6-*a*) for all calculated parameters in the present chapter, and the least squares linear fit gives

$$R_B \times V_B = (1.34646 \pm 2.3675 \times 10^{-4}) \times Z_T Z_P. \qquad (1.13)$$

The accuracy of the above formula is tested in figure (1.6-*b*) by comparing the product of $R_B\, V_B$ calculated using DFM and the values calculated using formula (1.13). The differences shown in figure (1.6-*b*) do not exceed 2%, except for a number of reactions between $^4He$ and some nuclei, reach to about 3%. In general, the accuracy of this formula is higher than 98% for all heavy and super heavy ion reactions. The above relation can efficiently used to calculate the potential barrier parameters, if an accurate method is valid to calculate one of them.

# *Chapter 2*

# *Universal function of nuclear proximity potential derived from M3Y nucleon-nucleon interaction*

# Chapter 2

# Universal function of nuclear proximity potential derived from M3Y nucleon-nucleon interaction

## 2.1 Introduction

Recently, the study of nuclear fusion has been of much interest, and the knowledge of the barrier parameters is also known to be important for fusion studies. Well knowledge of the barrier parameters (barrier height and position) helps to choose the optimum conditions for fusion to give the wanted results. The height of the barrier is especially important for the production of heavy and superheavy elements (SHE) [1]. In chapter (1), study of barrier parameters has been performed in the frame work of DFM, starting from empirical nuclear densities and M3Y nucleon-nucleon (NN) force. Systematic behavior of the fusion barrier parameters has been presented for large number of interacting ion pairs, and we discussed analytical expressions for calculation of the interaction barrier directly, starting from masses, charges, and/or radii of interacting nuclei.

Evaluation of nucleus-nucleus potential is a complicated problem, especially when the ground states of interacting nuclei are deformed [2- 4]. The nuclear part of a nucleus-nucleus potential has been studied in the framework of various models [5- 15]. The so-called “proximity” model [12, 15, 16] can easily and successfully be used to calculate the nuclear interaction between two medium-heavy or heavy nuclei. The original proximity interaction derived by Blocki et al [16] was obtained from a description of the interacting nuclei by the Thomas-Fermi approximation of the energy density of the ion-ion system. The expression of proximity interaction for spherical nuclei is given as

$$V_P(s_0) = 4\pi\bar{R}b\gamma\Phi(s_0) \tag{2.1}$$

The nuclear proximity potential, given above, is the product of two factors, one depending on the shape and geometry of the two nuclei ($\bar{R}b\gamma$), and the other is the universal function depending only on the shortest separation distance between them ($\Phi(s_0)$). In the first factor, $b$ is the diffuseness of the nuclear surface, given by

$$b = \frac{\pi}{2\sqrt{3}\, ln\, 9} \times t_{10-90} \qquad (2.2);$$

where $t_{10-90}$ is the thickness of the surface in which the nuclear matter density profile changes from 90% to 10%. For a 2pF distribution, $t_{10-90}$ is related to the diffuseness parameter *a* through: $t_{10-90} = 2a\, ln\, 9$, this give,

$$b = \frac{\pi}{\sqrt{3}} \times a$$

Many authors [16- 19] use a constant value of 1 *fm* or 0.99 *fm* as an approximate value of $b$ for heavy ion reactions. Any way, we will use the formula (2.2) to calculate the surface width (*b*) for any system, whatever the shape of density distribution. The diffused surface width for any reaction pair is taken as the mathematical average of the values obtained for each nucleus individually. $\gamma$ is the specific nuclear surface energy, given by

$$\gamma = 0.9517 \left[1 - 1.7826 \left(\frac{N-Z}{A}\right)^2\right] MeV\, fm^{-2} \qquad (2.3);$$

where N, Z and A are respectively the neutron, charge, and mass numbers of the bulk reaction system, *i.e.* for the reaction between two ions donated by 1 and 2,

$$N = N_1 + N_2$$
$$Z = Z_1 + Z_2\ ;$$
$$A = A_1 + A_2$$

$\bar{R}$ is the mean curvature radius of the reaction pair, given by

$$\bar{R} = \frac{R_1 R_2}{R_1 + R_2} \qquad (2.4);$$

where $R_1$ and $R_2$ are the radii of the two interacting nuclei.

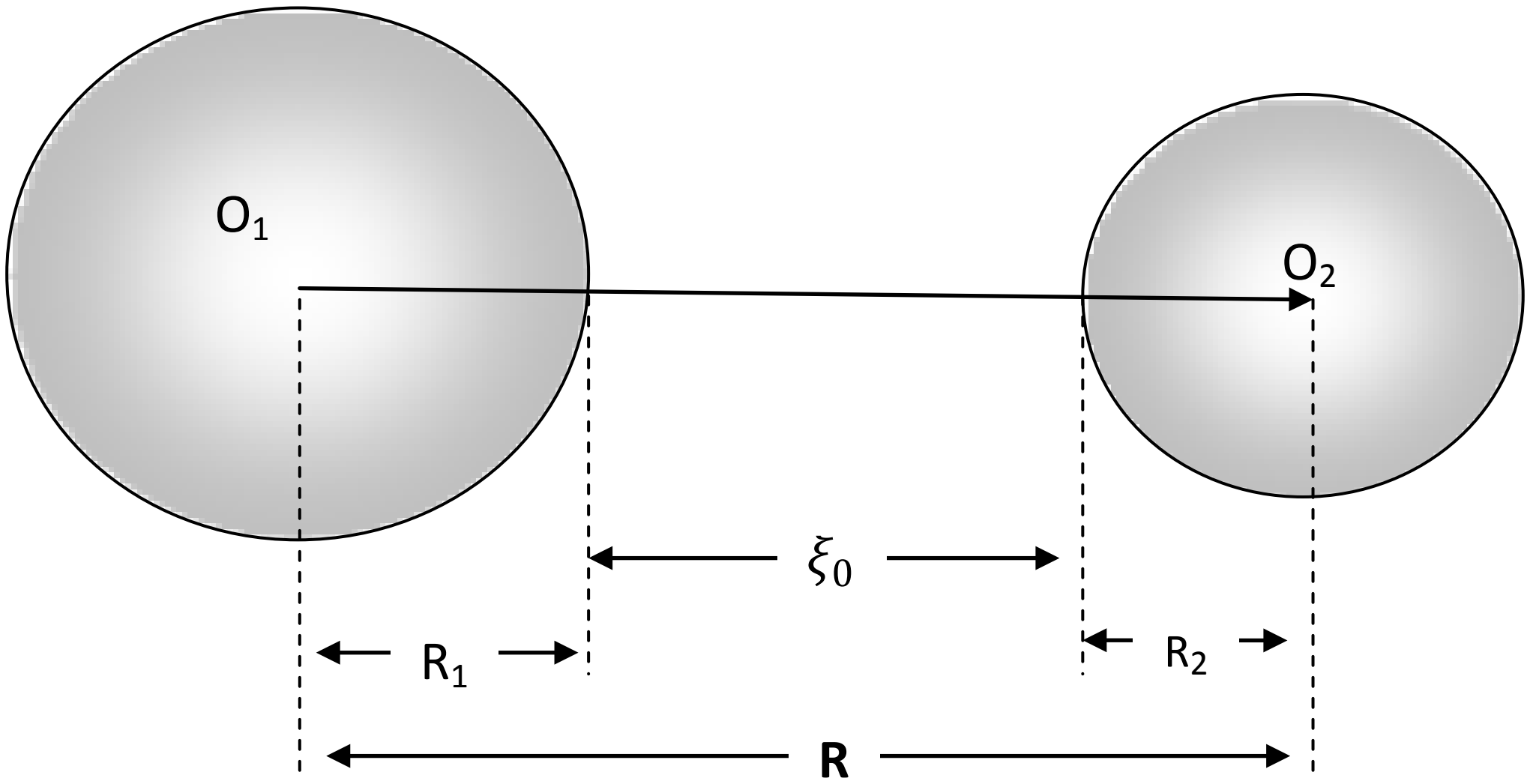

Figure (2.1)

The minimum separation distance between the nuclear surfaces is the essential quantity of the idea of proximity [16]. Figure (2.1) shows the minimum separation distance ( $\xi_0$ ) between the nuclear surfaces for interaction between two spherical nuclei, the dimensionless minimum separation distance ($s_0$) is the minimum separation distance ($\xi_0$) defined in units of $b$, i.e. $s_0 = \xi_0/b$. For the reactions between deformed and oriented nuclei the problem becomes complicated and calculation of $s_0$ can be done by using sophisticated iterative procedures just like introduced in references [12, 15].

The second factor in Equation (2.1) is the universal function $\Phi(s_0)$, independent of the shapes of nuclei or the geometry of nuclear system, but depends on the minimum separation distance $s_0$; it is therefore one function for all nuclei. The main advantage of the proximity potential is in the idea of universality (or system-independence). Authors in reference [16] (and references there) gave a rough approximation of the universal function $\Phi(s_0)$ by the following "cubic-exponential" pocket formula

$$\Phi(s_0) = \begin{cases} -\frac{1}{2}(s_0 - 2.54)^2 - 0.0852(s_0 - 2.54)^3, & s_0 \leq 1.2511 \\ -3.437 \exp\left(-\frac{s_0}{0.75}\right), & s_0 \geq 1.2511 \end{cases} \quad \textbf{(2.5)}$$

The equations mentioned above give only average estimates for doubly closed shell spherical-spherical ion interactions. Variations caused by shell effects and including deformations have to be taken into account. In particular, when using the proximity potential for deformed nuclei, the minimum separation distance $s_0$ as well as the mean curvature radius $\bar{R}$, have to be carefully related to the orientations and shapes of the nuclei. Recently, many studies have been done to derive the proximity potential between two deformed and oriented nuclei [3, 12, 15].

For various applications it is useful to have a simple analytical representation of the universal function which enters in the nucleus-nucleus proximity potential calculation; and it is more useful if that analytical representation agrees with the exact values for large number of interactions. For this purpose, we will use the DFM calculations, which have good agreement with experimental data [20], to give approximate representation of the universal function. It should be noted that the universal function had been derived from HI optical potential calculated from the two colliding nuclear matter approach [21-23]. Recently another shape of the universal function had been obtained from the Skyrme nucleus–nucleus interaction in the semi-classical extended Thomas–Fermi (ETF) approach [24, 25].

In the present chapter we are looking for a universal function of nuclear proximity potential derived from M3Y NN interaction in the frame work of DFM, which already considered in chapter (1) of the present thesis.

Since the folding model has its greatest validity in the tail of the potential at which the Thomas–Fermi method breaks down, the calculation of the universal function will be obtained from the DFM as a standard.

The objective of the present work is to make the study of the fusion barrier parameters easier and does not need to complicated numerical calculations as those exist in DFM calculations. This is achieved by deriving a universal function from the DFM considered in the first chapter. The derived universal function must have its greatest validity around the barrier position. For all reactions considered in chapter (1), we calculated the values of $s_0$ corresponding to the coulomb barrier radii, and we found that the values of $s_0$ are positive and located between 1 and 4. Thus the success of the obtained universal function in this study depends on the unique value of $\Phi(s_0)$ in that region.

## 2.2 Calculation of universal function starting from M3Y-Reid NN interaction

It is shown in section (1.2) that the nucleus-nucleus potential is a function of the center of mass separation distance ($R$) between the two interacting nuclei, and for spherical-spherical interacting pair interaction, it can be easily calculated by the integral (1.6). For spherical-spherical pair, the minimum separation distance between surfaces of the two nuclei is the distance between the nuclear surfaces along the line connecting the mass centers ($R$), as shown in figure (2.1), so, the minimum separation distance $\xi_0$ is defined as

$$\xi_0(fm) = R(fm) - R_1(fm) - R_2(fm)$$

and then, the dimensionless quantity $s_0$ is

$$s_0 = \frac{\xi_0}{b} = \frac{R - R_1 - R_2}{b} \qquad (2.6)$$

The proximity potential $V_P(s_0)$ at any distance must be equal to the value calculated through the DFM with M3Y-interaction ($V_{M3Y}(R)$ ), and then

$$V_P(s_0) = V_{M3Y}(s_0\, b + R_1 + R_2) \qquad (2.7)$$

$V_{M3Y}$ in this equation is the nuclear part of nucleus-nucleus potential calculated in the frame work of DFM discussed in section (1.2), the effective nucleon-nucleon interaction used is the well-known M3Y-Reid force [26, 27] in the form given by equation (1.7). The universal function for M3Y NN interaction can be obtained by using the definition of proximity potential given by equation (2.7) in the original proximity formula (2.1), this give the universal function as

$$\Phi(s_0) = \frac{V_{M3Y}(s_0\ b + R_1 + R_2)}{4\pi\bar{R}b\gamma} \tag{2.8}$$

Equation (2.8) defines the universal function for negative, zero, and positive values of $s_0$, however, the proximity model becomes less correct as the overlap of the two nuclei increases. This deviation may be a sign of more complex behavior of the nucleus–nucleus potential at small distances in the region of overlap of nuclear surfaces. For large overlaps the nuclear potential will be composed of an energy associated with the bulk of the overlap region, and of surface-layer energy. The proximity model considers only surface-layer interaction, but the DFM gives an average of the effective nucleon–nucleon potential over nuclear densities, and the effects due to nuclear-surface region are automatically taken into account. Therefore the universal function $\Phi(s_0)$, calculated from DFM results, can be approximated to one accurate expression only at the surface and beyond the overlap region at which the surface effect dominates. So, we extend our search for universal function for positive $s_0$ and around ($s_0$ = 0). In the following Figures we show the universal function calculated by equation (2.8) as a function of $s_0$, for reactions involving nuclei of mass numbers from 4 to 238. The results involved here are for symmetric reactions only, i.e., reactions between two identical ions. The reason of this choice will be discussed in details in the next section. Reactions involving $^{4}He$ will be discussed separately, because of its special effect on the calculated universal function. In the present chapter, we will introduce a universal function which is useful for barrier calculation for heavy ion reactions, and a similar universal function for reactions involving $^{4}He$.

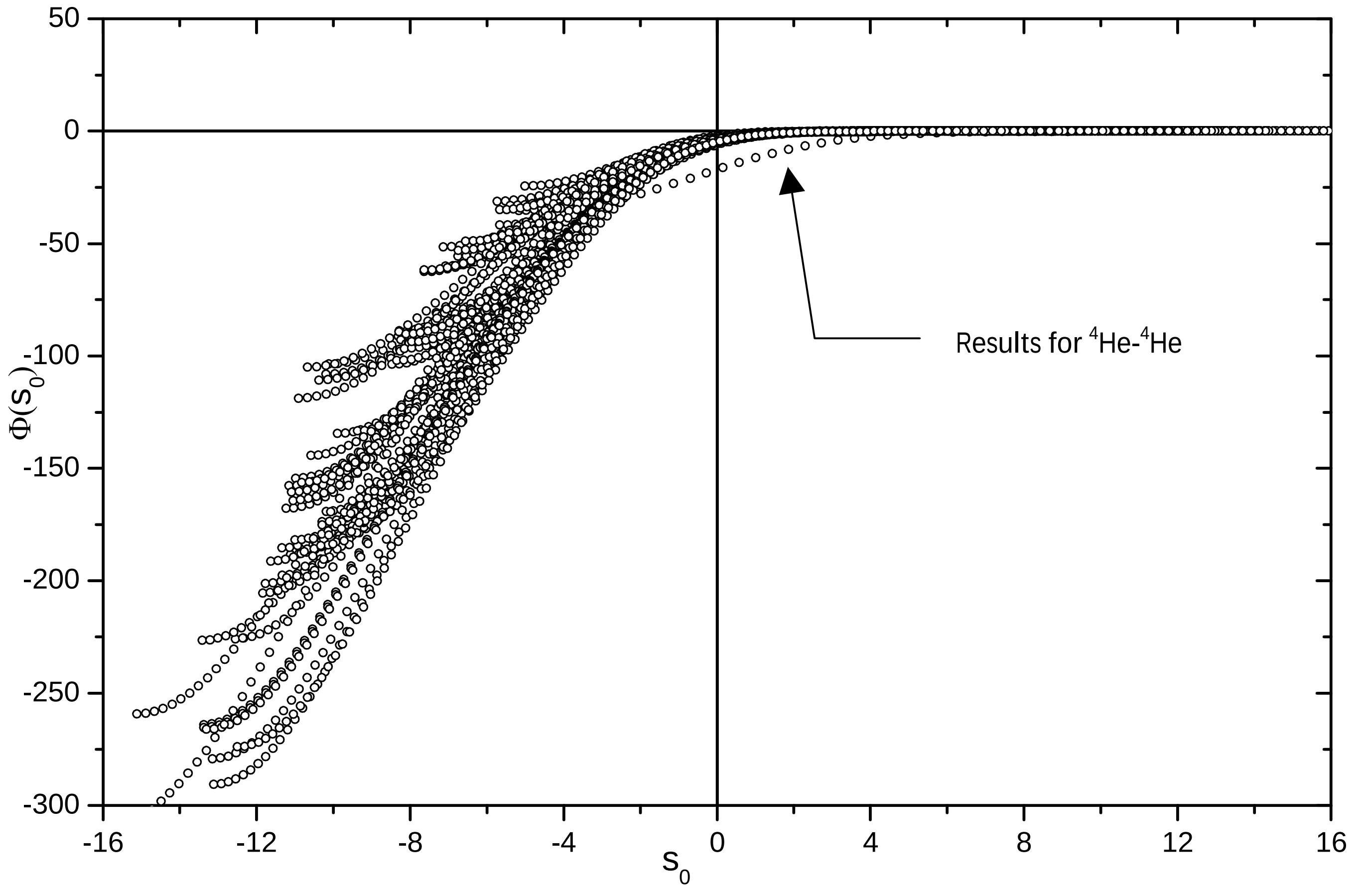

Figure(2.2-*a*) Universal function $\Phi(s_0)$ calculated from DFM with M3Y interaction as a function of the dimensionless separation $s_0$, for symmetric reactions between ions of mass numbers up to 238.

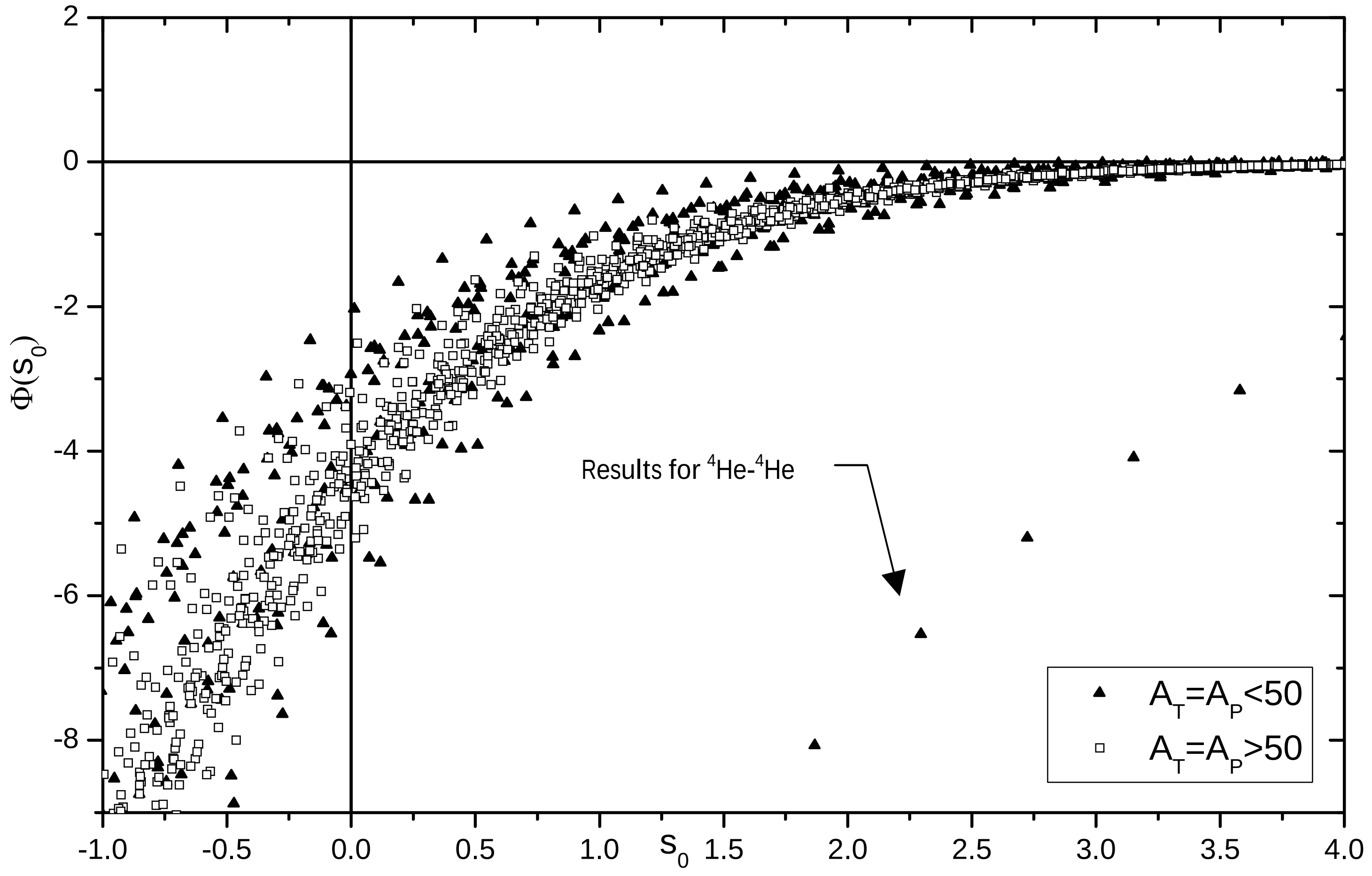

Figure (2.2-*b*) The same as Figure (2.2-*a*) but with $s_0$ varying in the range [-1, 4]

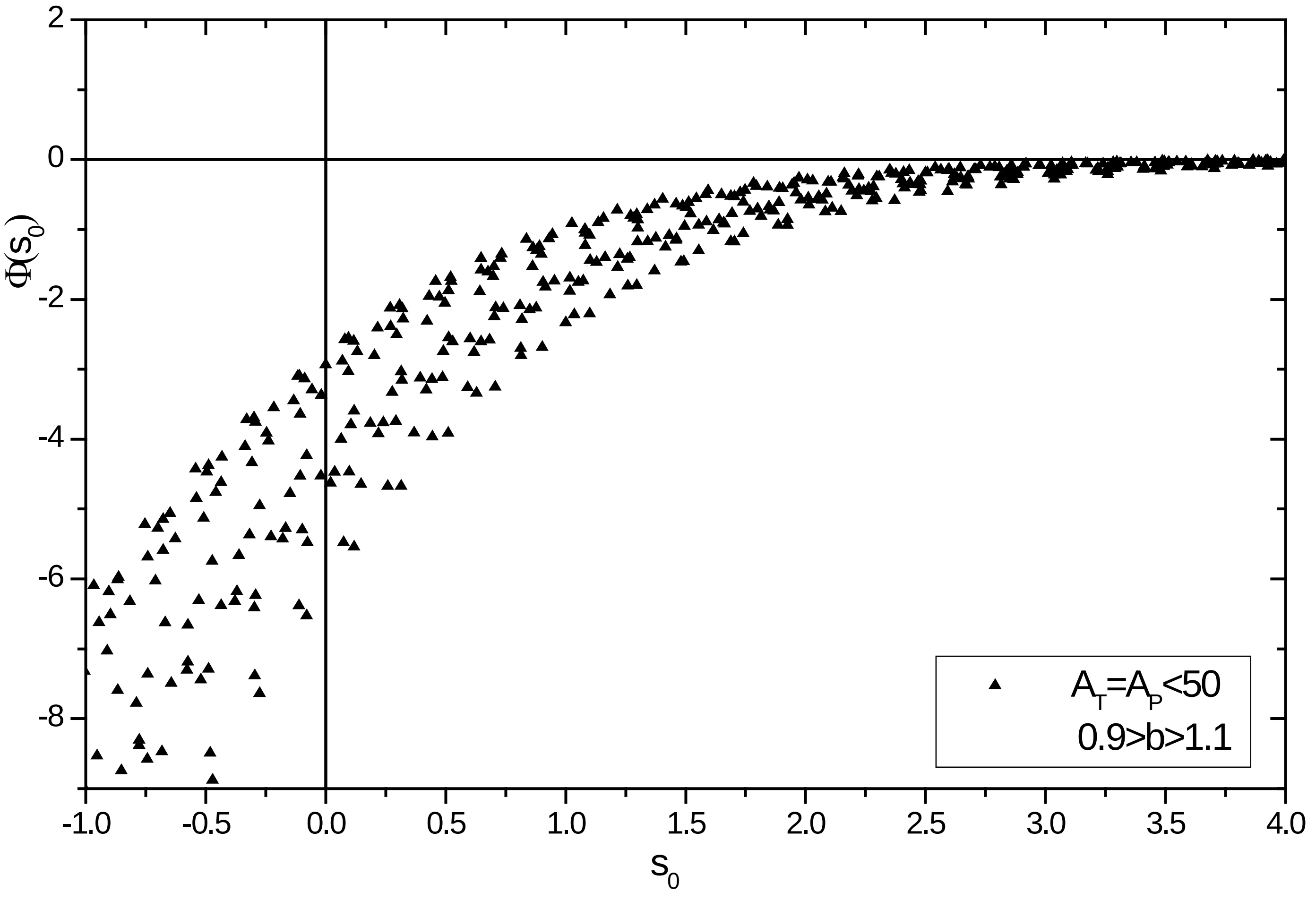

Figure (2.3-*a*) Calculated universal function $\Phi(s_0)$ as a function of the dimensionless separation $s_0$, for reactions involve ions of mass numbers up to 50,and surface diffuseness between 0.9 and 1.1.

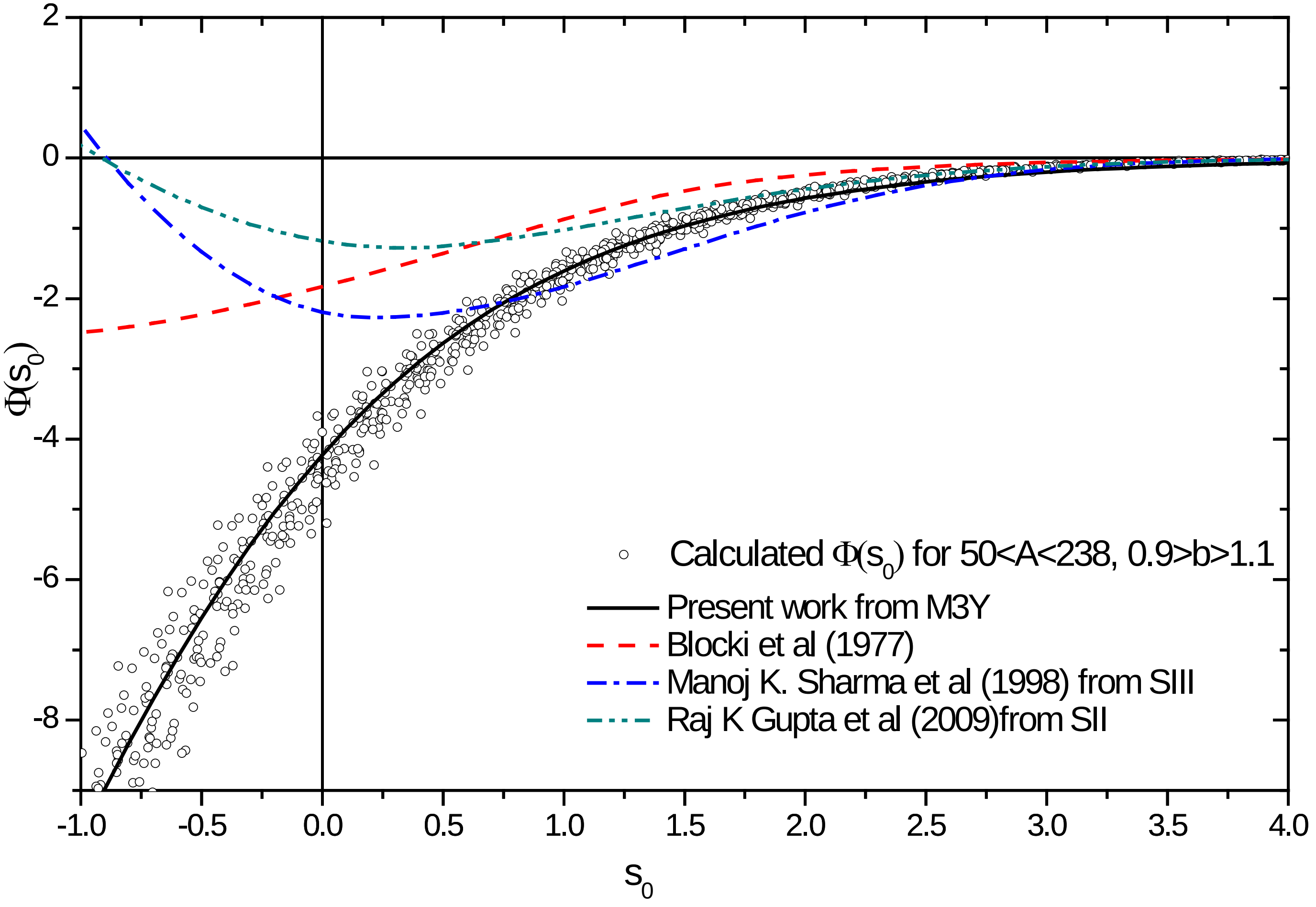

Figure (2.3-*b*) The same as Figure (2.3-*a*) but for reactions involve ions with mass numbers *A* as: 50>A>238. Different shapes of the universal function are plotted [16, 24, 25].

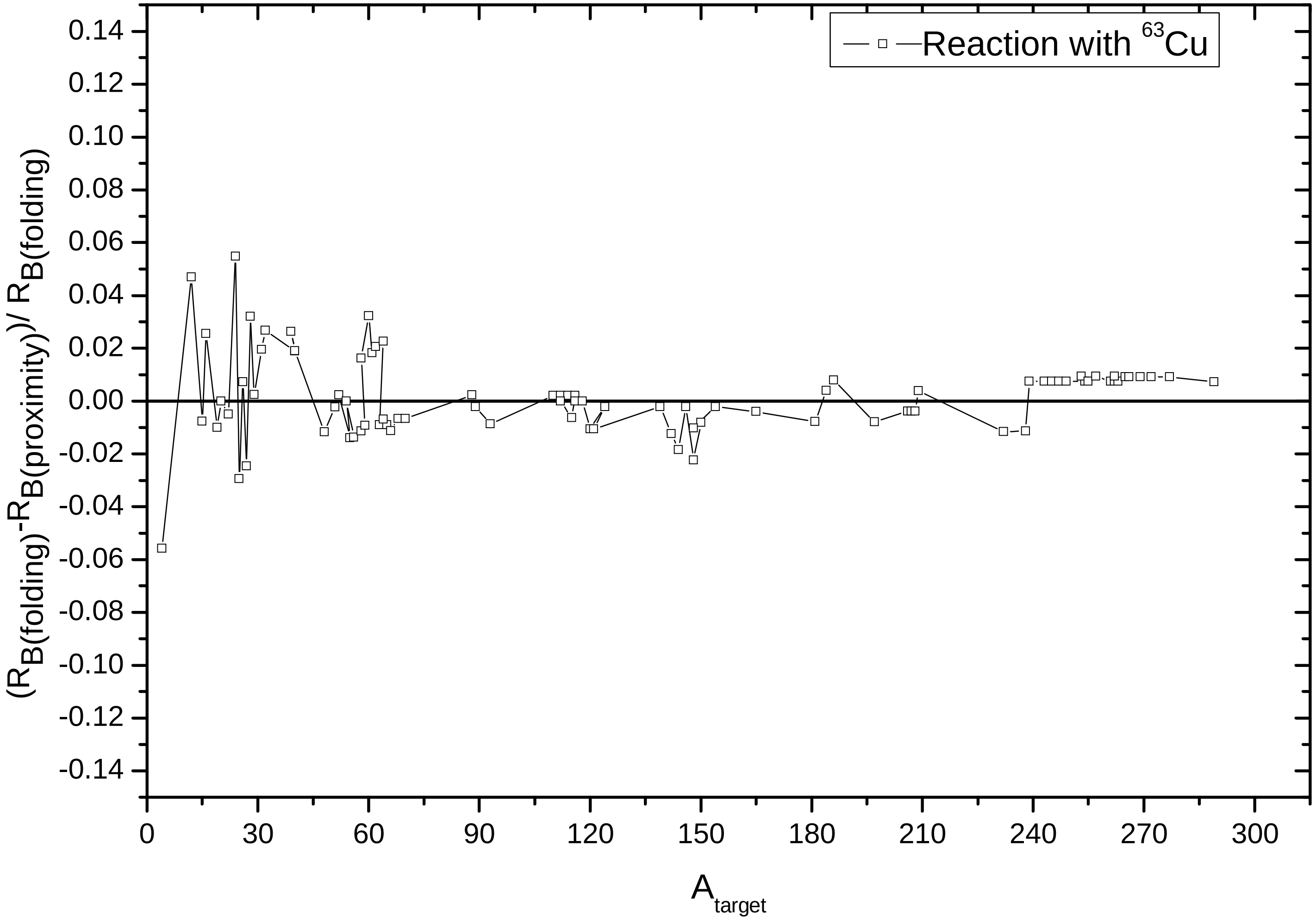

Figure (2.4-*a*) Fractional error between the values of potential barrier position calculated using the proximity model with universal function given by formula (2.9), and values calculated using DFM, for reactions between $^{63}Cu$ and different ions.

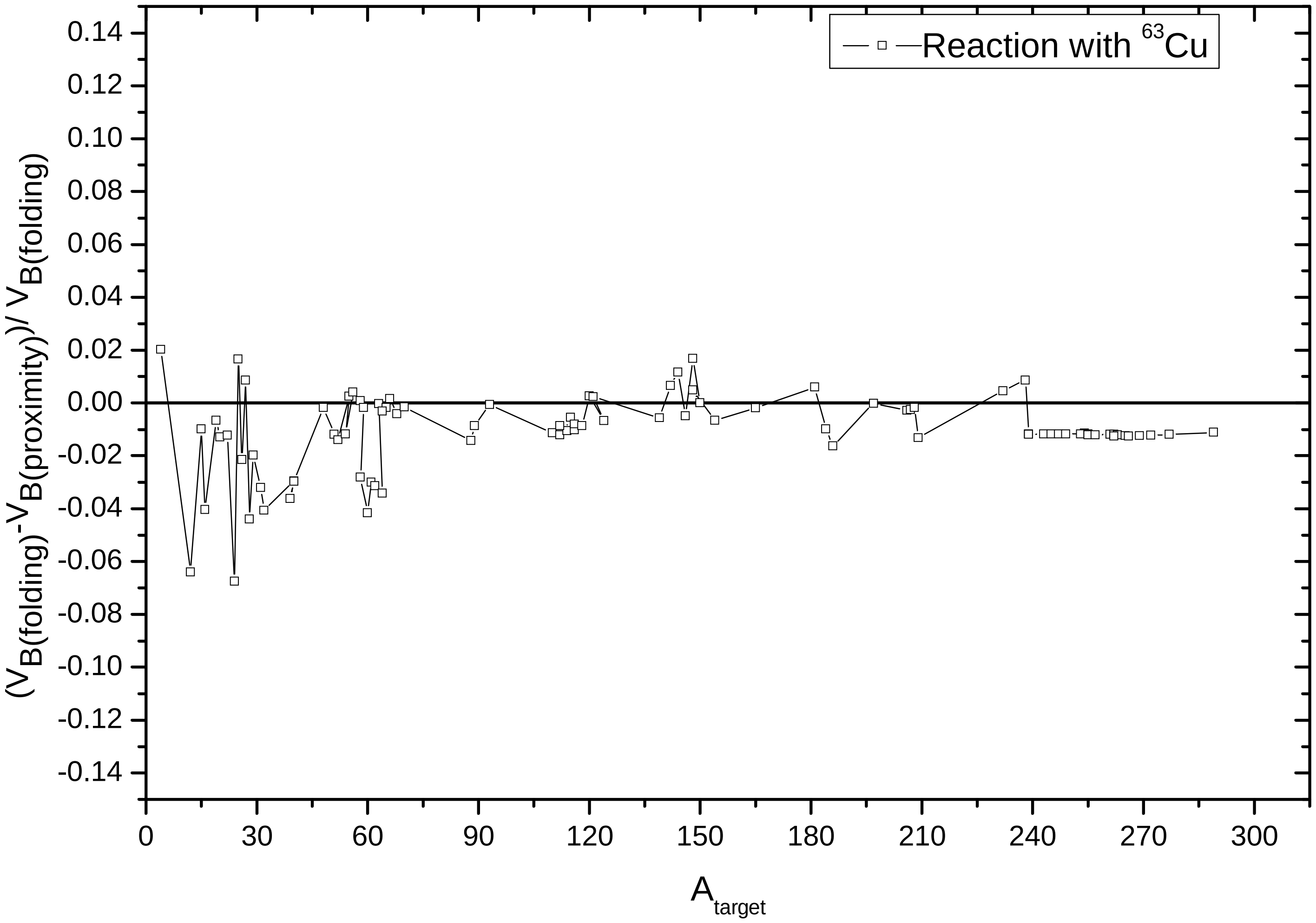

Figure (2.4-*b*) Fractional error between the values of potential barrier height calculated using the proximity model with universal function given by formula (2.9), and values calculated using DFM, for reactions between $^{63}Cu$ and different ions.

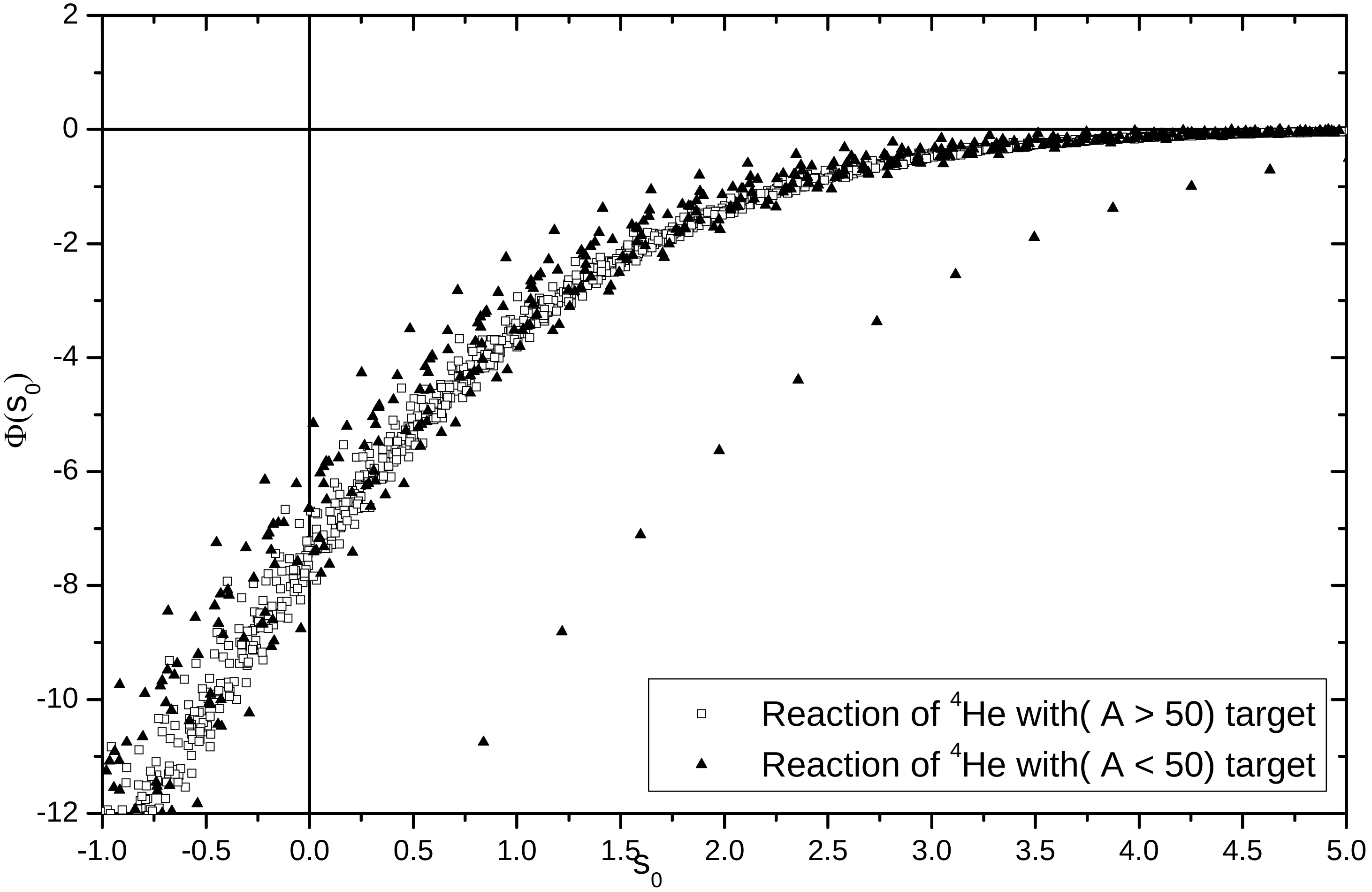

Figure (2.5-*a*) Calculated universal function $\Phi(s_0)$ as a function of the dimensionless separation $s_0$, for reactions between $^4He$ and target nuclei with mass numbers up to 238.

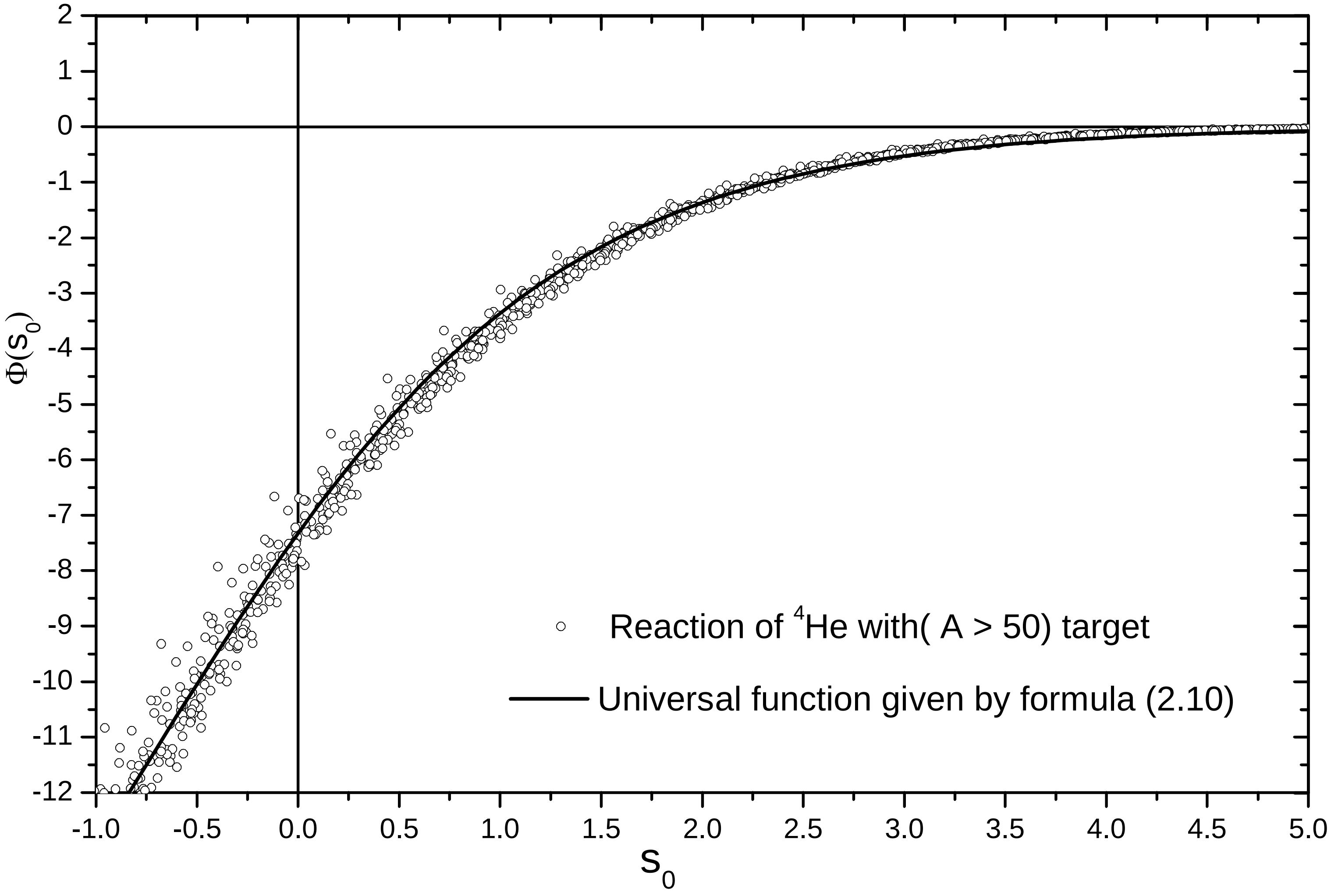

Figure (2.6-*b*) Same as Figure (2.5-*a*), but for target nuclei of mass numbers A as, 50>A>238.The solid line is the universal function derived from the best fit to data, given by formula (2.10).

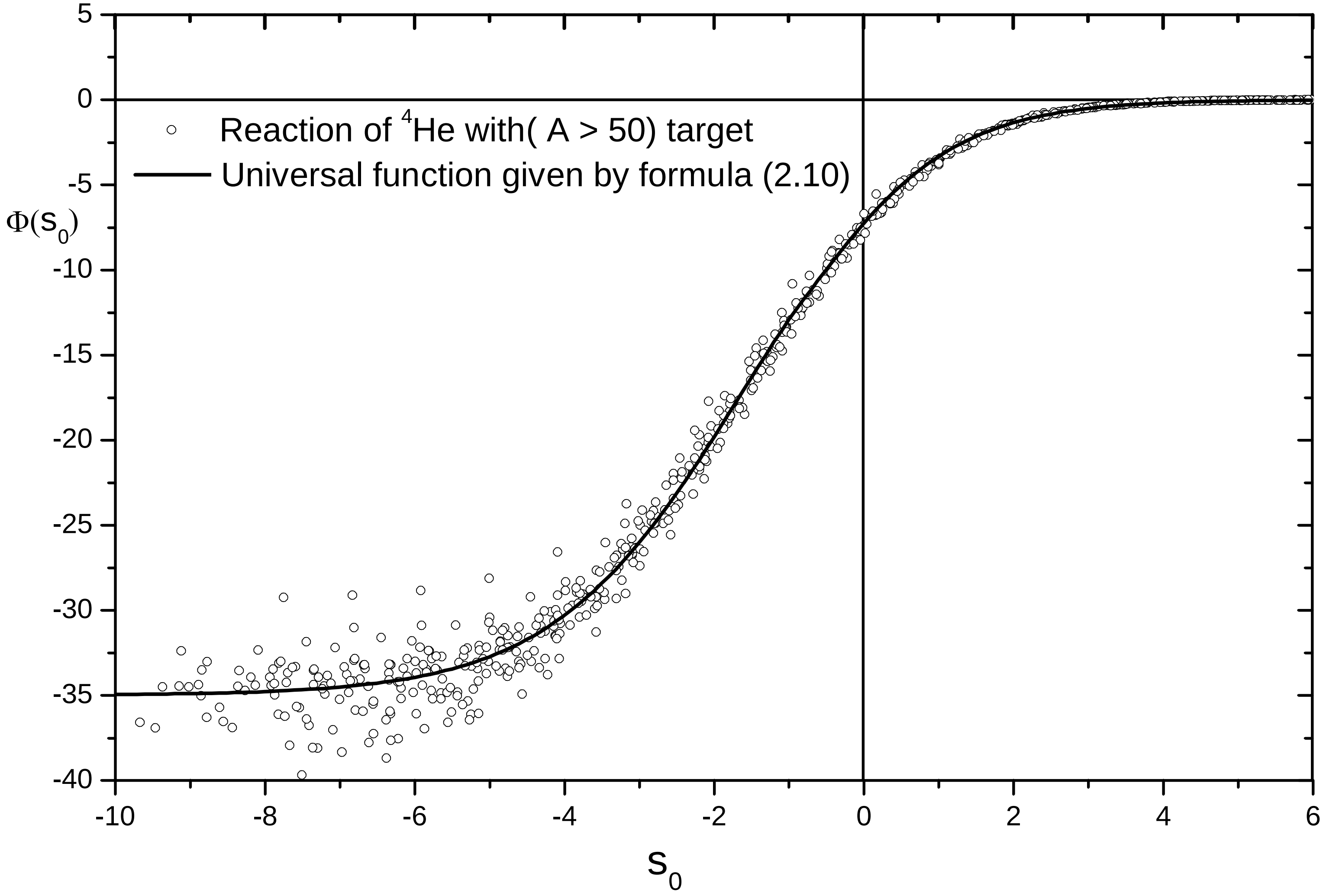

Figure (2.5-*c*) Same as Figure (2.5-*b*), but $s_0$ range varying in the range [-10,6]

## 2.3 Discussion

In the present chapter we use a method to drive a realistic universal function of proximity approach starting from realistic NN potential and nuclear density. For this purpose, we study only, the reactions involving ions which have density distributions given by the empirical formulae [28] listed in table (1.1). The idea of universality tells us that the values of the calculated universal function for all interactions must be sharply localized around an average line as a function of $s_0$. From the calculations done in the present chapter, it is found that some ions like $^{4}He$ and $^{60}Ni$ are far from the average $\Phi(s_0)$ line. Both the target and the projectile affect the value of the calculated universal function and the final result shows the average effect of the reaction pair. For example, considering an ideal symmetric pair, that gives exactly the average $\Phi(s_0)$ with no deviation, if one ion of this pair is replaced by another ion which causes an increase/decrease in the calculated universal function, the obtained line will be above/ below the average line. Moreover, if each of interacting ions leads to rising/lowering in the calculated universal function, the obtained universal function line will be more raised/ lowered. On this basis, symmetric reaction enhances the effect of any nucleus on the calculated universal function, because the reaction partners have exactly the same effect. We consider the results of $\Phi(s_0)$ derived from 70 interacting pair. The results involved here are for symmetric reactions only, starting from $^4He$-$^4He$ up to $^{238}U$- $^{238}U$. The studied 70 reactions can efficiently give a complete picture of the possible 2485 reactions occurring between different isotopes in the range of study. It is worth mentioning that the main objective of the present work is to study the fusion barrier, and then we will give more attention to the data points around the barrier position ($-1< s_0< 5$).

In figure (2.2-*a*) the universal function $\Phi(s_0)$ calculated from M3Y interaction is plotted as a function of the dimensionless separation $s_0$, for symmetric reactions between ions of mass numbers up to 238. It is obvious that the dispersion of different sets of data points increases as $s_0$ become more negative. The proximity model gives average estimate of the nuclear potential considering the surface to surface gap and the surface energy density. The DFM, unlike the proximity model, gives an average over all interactions, and the surface effect is automatically involved in the folded potential. For positive separations, largest contribution of the folding potential comes from surface interaction, and $\Phi(s_0)$, which stems from surface interaction, tends to be the same for all reactions.

In figure (2.2-*b*), the values of $\Phi(s_0)$ is plotted around the touching point and up to values which expected to be around the barrier positions. This figure shows that the data points for reactions involving nuclei of mass numbers less than 50 are more scattered than the rest of studied reactions. The spread of points in figure (2.2-*b*) indicates that it is impossible to get reasonable average universal function from M3Y NN force all over the range of the potential. The reaction between two $^{4}He$ nuclei has diffuseness ($b$) of 0.587 *fm*. while other reactions have diffuseness ($b$) around 1 *fm*. Moreover, $^{4}He$ nucleus has central nuclear density of 0.4229 $nucleon.fm^{-3}$, while the rest of nuclei have central nuclear density of about 0.17 $nucleon.fm^{-3}$. And it is also obvious the odd result for the reactions involving $^{4}He$, this is must be considered as a special case because of the unique nature of $^{4}He$. The reactions between $^{4}He$ projectile and other nuclei will be discussed separately later in this section.

Data included in figure (2.3-*a*, *b*) are for reactions with diffuseness ($b$) has values between 0.9 *fm* and 1.1 *fm*, and other reactions are excluded. It is found that some reactions which have smaller or higher values of $b$ may increase the scattering of data points, for example, $^{60}Ni - {}^{60}Ni$ reaction has diffuseness $b$ = 0.842 *fm*, and at $s_0 = 0$, $\Phi(s_0)$ is higher by about 0.5 from the nearest data point for all pairs including ions of mass numbers as $50 \leq A \leq 238$. It is worth mentioning that there is only little number of nuclei which excluded; however, these nuclei will give more accurate results for their reactions with other nuclei.

The universal function of proximity is expected -from the original definition of proximity model- to be diffuseness independent, but the effect observed here is coming from the bulk nuclear reactions and not the surface-surface interaction. The nucleus is not sharp edged, but it has a diffused surface, so the half density radius -in some models- is taken to be the radius of the nucleus; and then there are a number of nucleons beyond the radius of the nucleus. In a given reaction between two nuclei, at $s_0 = 0$, the nucleons in the tail (nucleons beyond the radius of the nucleus) of each nucleus are completely inside the other, and the value of potential must be lower than calculated considering only the surface effect. Considering this idea, it is expected that reactions with higher diffuseness must have more negative $\Phi(s_0)$ at the same separations. The most of nuclei have diffuseness parameters within a small range around an average value, which make the differences due to tail nucleons overlap tend to be the same. It is clear from figure (2.3-*b*) that the values of the universal function at zero $s_{0,}$ spread around -4.5 within a range of about 1.8, and the spread of data around an average value decreases as $s_0$ increases. So, it is impossible to state a formula for the universal function as a function of just one variable $s_0$ with accuracy higher than a certain limit. Inserting other parameters, such as the diffuseness parameter, into the universal function may improve the accuracy of the analytical formula produced, but in this case the name "universal" become no longer suitable for the new function, because it will lose its greatest property which is its universality.

The analytical formula of the universal function suitable for the present work has two main properties, the first: it has its greatest accuracy around the barrier position; the second: it must be a function of one variable $s_0$. In other words, we have to give up some of accuracy to keep the universality of the universal function. This formula can be obtained from the best fit to the data around the barrier position. For this purpose we fit all the data points for $s_0$ larger than -1in figure (2.3-*b*), the least square fit gives

$$\Phi(s_0) = \frac{-30}{1 + e^{\frac{s_0+1.57}{0.92}}}, \qquad s_0 > -1.0 \qquad (2.9)$$

This formula can easily used in equation (2.1) to give the nuclear interaction between two heavy ions, then adding the Coulomb potential we obtain the Coulomb barrier. Different shapes of universal function [16, 24, 25] are plotted in figure (2.3-$b$), the common feature between the universal functions derived in these references is the presence of hard core, this effect can appear if we only consider the surface effects as in reference [16], or if the universal function is derived from a potential having the same shape. The universal function introduced in the present chapter does not contain a hard core like that obtained from Skyrme forces [24, 25], but it has the same shape as M3Y nucleus-nucleus potential. To test the ability of formula (2.9) in producing the parameters of the Coulomb barrier, we chose the reactions of $^{63}Cu$ as a fixed reactant with all ions given in table (1.1) beside its reactions with the trans-uranium elements with densities parameterized as presented in chapter (1). Figures (2.4-$a$, $b$) show the relative deviations between the results of DFM and the results derived from equation (2.9) for barrier position ($R_B$) and barrier height ($V_B$) respectively. In both figures, the reactions with light nuclei show large deviations, but for heavy and super heavy ions the deviations never exceed the 2%. However the shape of universal function given by equation (2.9) is useful for barrier calculation, it can not give a complete description of the nuclear part of heavy ion interaction.

The calculated universal function for reactions involve $^{4}He$ projectile is illustrated in figures (2.5), it is obvious that the data points are less scattered than in figures (2.3), this is because of presence of fixed projectile, i.e. the deviations of different data sets comes from the effect of only one nucleus (the target). Fitting data points around the barrier position gives a function similar to (2.9), but in this case the obtained function can describe the interaction between $^{4}He$ and heavy ions. Presence of such a universal function for $^{4}He$ potential is very important in the field of α-particle studies. Recently proximity approach have been used by many authors to study half life times of α-decay of heavy and super heavy ions assuming proximity approach for barrier penetration. The same approach has been used for α-potential inside the nucleus. Those studies require a well defined universal function for any separation distance ($s_0$). Averaging over the calculated $\Phi(s_0)$ for all data points can give a single analytical formula for $\Phi(s_0)$ for any $s_0$, but this formula become inaccurate for barrier calculations. To keep the accuracy for any $s_0$, we fitted data around the barrier position separately, then the best formulation of $\Phi(s_0)$ for the reaction between $^{4}He$ and HI, is

$$\Phi_{He}(s_0) = \begin{cases} \dfrac{-30}{1 + e^{\frac{s_0+1.17}{1.1}}}, & s_0 \geq -0.943 \\ \dfrac{-35}{1 + e^{\frac{s_0+1.66}{1.25}}}, & s_0 \leq -0.943 \end{cases} \qquad (2.10)$$

which is useful for barrier calculation and give approximate value of the universal function for any separation distance.

Formula (2.10) can be used to approximate the nuclear potential for the reaction between $^{4}He$ and HI of mass numbers between 50 and 238.

# *Appendices*

# Appendix 1

# DFM and Fourier transformation

The Fourier transform of a function *f(**r**)* defined by

$$\tilde{f}(\boldsymbol{k}) = \int f(\boldsymbol{r})\, e^{(i\boldsymbol{k}.\boldsymbol{r})}\, d\boldsymbol{r}. \qquad (\boldsymbol{App.1-1})$$

so that

$$f(\boldsymbol{r}) = (2\pi)^{-3} \int \tilde{f}(\boldsymbol{k})\, e^{(-i\boldsymbol{k}.\boldsymbol{r})}\, d\boldsymbol{k}. \qquad (\boldsymbol{App.1-2})$$

The multipole expansion of the exponential term in (App.1-1) is given by

$$e^{(i\boldsymbol{k}.\boldsymbol{r})} = 4\pi \sum_{l,m} i^l\, j_l\,(kr) Y_{lm}(\Omega_{\boldsymbol{r}})\, Y^*_{lm}(\Omega_{\boldsymbol{k}}).$$

so that

$$\tilde{f}(\boldsymbol{k}) = 4\pi \sum_{l,m} i^l\;\; Y^*_{lm}(\boldsymbol{\Omega_k}) \int j_l(kr)\, f(\boldsymbol{r})\, r^2\, d\boldsymbol{r}\, Y_{lm}(\Omega_r)\, d\Omega_r.$$

From the properties of spherical harmonics following

$$\int\; Y_{lm}(\Omega)\, d\Omega\; = \sqrt{4\pi}\, \boldsymbol{\delta_{l0}}\, \boldsymbol{\delta_{m0}}.$$

$$Y_{00}(\Omega) = \frac{1}{\sqrt{4\pi}}.$$

and if $f(\boldsymbol{r})$ in (App.1-1) is a scalar function of $r$ and**,** then $\tilde{f}(k)$ has the form

$$\tilde{f}(k) = 4\pi \int j_0(kr)\, f(r)\, r^2\, dr. \qquad (\boldsymbol{App.1-3})$$

Similarly, the back Fourier transformation becomes

$$f(r) = 4\pi(2\pi)^{-3} \int j_0(kr)\, \tilde{f}(k)\, k^2\, dk. \qquad (\boldsymbol{App.1-4})$$

The expression of $\tilde{f}(k)$ given by (App.1-3) is very useful in the calculation of interaction potentials Fourier transform, as will be shown below.

## Appendix 1-A: DFM in momentum space

The Fourier transform of the folded quantity is simply the product of the transforms of the individual component functions, as convolution theorem states. This makes the calculation much easier than directly doing the folding integrals. Making use of convolution theorem the DFM integral (App.1-3)

$$U(\boldsymbol{R}) = \int d\boldsymbol{r_1} \int d\boldsymbol{r_2}\ \rho_1(\boldsymbol{r_1}) v(\boldsymbol{r_{12}}) \rho_2(\boldsymbol{r_2}). \qquad (\boldsymbol{App.1-5})$$

$$\widetilde{U}(\boldsymbol{k}) = \tilde{\rho}_1(\boldsymbol{k}) \tilde{v}(\boldsymbol{k}) \tilde{\rho}_2(-\boldsymbol{k}). \qquad (\boldsymbol{App.1-6})$$

If the nuclear density distribution $\rho_i(k)$ is spherically symmetric, we can apply the simplified expression (App.1-3) to get

$$\tilde{\rho}_i(k) = 4\pi \int j_0(kr)\, \rho_i(r)\, r^2\, dr$$

So that

$$\widetilde{U}(\boldsymbol{k}) = \left[4\pi \int j_0(-kr_1)\, \rho_1(r_1)\, r_1^2\, dr_1\right] \tilde{v}(\boldsymbol{k}) \left[4\pi \int j_0(-kr_2)\, \rho_2(r_2) r_2^2\, dr_2\right]$$

$$(\boldsymbol{App.1-7})$$

Applying (App.1-2) on the total potential

$$U(\boldsymbol{R}) = (2\pi)^{-3} \int \widetilde{U}(\boldsymbol{k})\, e^{(-i\boldsymbol{k.R})}\, d\boldsymbol{k}.$$

$\widetilde{U}(\boldsymbol{k})$ is a scalar function as $\tilde{v}(\boldsymbol{k})$ is, then applying (App.1-4)

$$U(R) = 4\pi(2\pi)^{-3} \int k^2 J_0(kR)\, \widetilde{U}(k)\ dk$$

$\widetilde{U}(k)$ is defined in(App.1-7), then

$$U(R) = \left(\frac{1}{2\,\pi^2}\right)\int k^2\,J_0(kR)\,\left[4\pi\int j_0(kr_1)\,\rho_1(r_1)\,r_1^2\,dr_1\right]\tilde{v}(k)\left[4\pi\int j_0(-kr_2)\,\rho_2(r_2)r_2^2\,dr_2\right]\,dk$$

which finally simplified to

$$U(R) = 8\int k^2\,J_0(kR)\,\tilde{v}(k)\left[\int j_0(kr_1)\rho_1(r_1)r_1^2\,dr_1\right]\left[\int j_0(kr_2)\,\rho_2(r_2)r_2^2 dr_2\right]dk \qquad (\mathbf{App.1-8})$$

The expression of $U(R)$ given by (App.1- 8) is a powerful in the calculation of interaction potential between two spherical nuclei. Moreover, it is easy to calculate even numerically.

# Appendix 1-B: Fourier transform of coulomb and M3Y potentials

## B.1- Fourier transform of coulomb potentials

The coulomb potential between two protons of charge $e$ and separated by distance $r$ is

$$v_c(\boldsymbol{r}) = \frac{k_e e^2}{r}$$

Applying Fourier transformation given by (App.1- 3) for $v_c(r)$

$$\tilde{v}_c(k) = 4\pi k_e e^2 \int \frac{\sin(kr)}{kr} \frac{1}{r} r^2 \, dr$$

$$\tilde{v}_c(k) = \frac{4\pi k_e e^2}{k} \int \sin(kr) \, dr$$

replacing $\sin(kr)$with its exponential form

$$\sin(x) = \frac{e^{ix} - e^{-ix}}{2i}$$

then we have

$$\tilde{v}_c(k) = \frac{4\pi k_e e^2}{k} \int \frac{e^{ikr} - e^{-ikr}}{2i} dr$$

making use of Yukawa trick

$$\tilde{v}_c(k) = \lim_{a \to 0} \frac{4\pi k_e e^2}{k} (e^{-ar}) \int \frac{e^{ikr} - e^{-ikr}}{2i} dr$$

$$\tilde{v}_c(k) = \lim_{a \to 0} \frac{4\pi k_e e^2}{k} \int \frac{e^{-(a-ik)r} - e^{-(a+ik)r}}{2i} dr$$

$$\tilde{v}_c(k) = \lim_{a \to 0} \frac{4\pi k_e e^2}{2ik} \left[ \frac{e^{-(a-ik)r}}{-(a-ik)} + \frac{e^{-(a+ik)r}}{a+ik} \right]_0^\infty$$

$$\tilde{v}_c(k) = \lim_{a \to 0} \frac{4\pi k_e e^2}{2ik} \left[ \frac{1}{(a-ik)} - \frac{1}{a+ik} \right]$$

$$\tilde{v}_c(k) = \frac{4\pi k_e e^2}{2ik}\left[\frac{2i}{k}\right]$$

so that, the coulomb potential in momentum-space is

$$\tilde{v}_c(k) = \frac{4\pi k_e e^2}{k^2} \qquad (\boldsymbol{App.1-9})$$

**B.2- Fourier transform of M3Y potentials**

The first two terms of M3Y interaction potential have the shape of Yukawa potential which has the form of

$$v_Y(\boldsymbol{r}) = \frac{V_0 e^{-\alpha r}}{\alpha r}.$$

Applying Fourier transformation given by (App.1- 3) for $v(r)$

$$\tilde{v}_Y(k) = 4\pi \int \frac{\sin(kr)}{kr} \frac{V_0 e^{-\alpha r}}{\alpha r} r^2 \, dr$$

$$\tilde{v}(k) = 4\pi \frac{V_0}{k\alpha} \int \sin(kr) \; e^{-\alpha r} \; dr$$

replacing $\sin(kr)$with its exponential form

$$\tilde{v}_Y(k) = \frac{4\pi V_0}{2ik\alpha} \int e^{(ik-\alpha)r} - e^{-(ik+\alpha)r} \; dr$$

$$\tilde{v}_Y(k) = \frac{4\pi V_0}{2ik\alpha} \left[\frac{e^{-(-ik+\alpha)r}}{ik-\alpha} + \frac{e^{-(ik+\alpha)r}}{ik+\alpha}\right]_0^\infty$$

$$\tilde{v}_Y(k) = \frac{4\pi V_0}{2ik\alpha}\left[\frac{1}{-ik+\alpha} - \frac{1}{ik+\alpha}\right]$$

$$\tilde{v}_Y(k) = \frac{4\pi V_0}{2ik\alpha}\left[\frac{ik+\alpha+ik-\alpha}{k^2+\alpha^2}\right]$$

$$\tilde{v}_Y(k) = \frac{4\pi V_0}{2ik\alpha}\left[\frac{2ik}{k^2+\alpha^2}\right]$$

$$\tilde{v}_Y(k) = \frac{4\pi V_0}{\alpha[k^2+\alpha^2]} \qquad \textbf{(App.1-10)}$$

The exchange term in M3Y potential has the form of delta shape potential of the form

$$v_D(\boldsymbol{r}) = V_0 \boldsymbol{\delta}(r)$$

using the integral

$$\int e^{(-i\boldsymbol{k}.\boldsymbol{r})}\, d\boldsymbol{k} = (2\pi)^3 \delta(\boldsymbol{r})$$

and comparing with the expression of Fourier transformation (App.1-2), therefore, the function ($f(\boldsymbol{k}) = 1$) is the Fourier transform of $\left(f(\boldsymbol{r}) = \delta(\boldsymbol{r})\right)$, and the exchange term in momentum space becomes

$$\tilde{v}_D(k) = V_0 \qquad \textbf{(App.1-11)}$$

M3Y potential that used in this work has the shape of

$$v_n(r) = \frac{7999\, e^{-4r}}{4r} - \frac{2134\, e^{-2.5r}}{2.5r} - 262\boldsymbol{\delta}(r)$$

Making use of (App.1-10) and (App.1-11), so that, the M3Y potential in momentum-space is

$$\tilde{v}_n(k) = \frac{7999 \times 4\pi}{4[k^2 + 4^2]} - \frac{2134 \times 4\pi}{2.5[k^2 + 2.5^2]} - 262 \qquad (\boldsymbol{App.1-12})$$

# Appendix 2

## Units of Coulomb constant

The coulomb potential between two protons of charge $e$ and separated by distance $r$ is

$$v_c(\boldsymbol{r}) = \frac{k_e e^2}{r}$$

and

$$k_e = \frac{1}{4\pi\varepsilon_0}$$

where the constant $\varepsilon_0$ is the *permittivity of free space* and has the value $8.8542 * 10^{-12}\mathrm{F/m}$ , and $e = 1.60219 * 10^{-19}$ is the charge of proton. So that, the potential obtained using these constants, with there SI units, is in Joules when the distance is in meters.

If the *fm* is used as a unit of *r* the permittivity of free space convert to

$$\varepsilon_0 = 8.854\ 2 * 10^{-27} \quad F/fm.$$

and the potential convert to be in MeV as

$$v_c(r) = \left(\frac{1}{4\pi\varepsilon_0}\frac{e^2}{r}\right)\left(\frac{10^{-6}}{e}\right) \quad \mathrm{MeV}$$

then, the Coulomb potential is given in MeV ,when *r* is given in fm, by

$$v_c(r) = (1.439\ 974\ 579)\left(\frac{1}{r}\right) \quad \mathrm{MeV} \qquad (\boldsymbol{App.2-1})$$

# المستخلص

فى الفصل الاول تم حساب بارامترات حاجز الجهد (مكان الحاجز وارتفاعه) لعدد كبير من أزواج الأنوية المختلفة باستخدام نموذج الطى المزدوج، بدءاً من دالة واقعية لتوزيع المادة النووية داخل الأنوية. وتمت دراسة سلوك بارامترات حاجز الجهد مع أعداد الكتلة، والشحنات، و أنصاف أقطار الأنوية المتفاعلة. تم تقديم السلوك المنتظم لبارامترات حاجز الجهد في صورة صيغ تحليلية بسيطة، يمكن استخدامها لحساب مكان وارتفاع حاجز الجهد مباشرة، كما تظهر العوامل التي يمكن أن تؤثر عليها.

في الفصل الثاني تم استخدام قيم الجهد الناتجة من حسابات نموذج الطى المزدوج لاستنتاج دالة عامة للتقارب النووي (universal function of nuclear proximity) يمكن استخدامها لحساب الجهد بين نواتين حول حاجز الجهد. كما تم استخدام الدالة الجديدة لحساب بارامترات حاجز الجهد ووجد ان الفرق بينها وبين القيم المحسوبة باستخدام نموذج الطى المزدوج لاتتجاوز (2%) لكل الأنوية الثقيلة والفائقة الثقل. كما تم تقديم دالة مماثلة لحساب الجهد بين الأنوية الثقيلة و جسيمات الفا.

# دراسة بارامترات حاجز الجهد بين أزواج الأنوية المختلفة

إعداد

**عبدالغنى رضا عبدالغنى أحمد**

رسالة مقدمة إلي

**كلية العلوم**

كجزء من متطلبات الحصول علي درجة

الماجستير  فى العلوم

( فيزياء نظرية )

قسم الفيزياء

كلية العلوم

جامعة القاهرة

( ٢٠١٠ )